\documentclass[pre,showpacs,showkeys,preprintnumbers,amsmath,amssymb,superscriptaddress,twocolumn]{revtex4-2}
\usepackage{graphicx}
\usepackage{amsfonts}
\usepackage{url}
\usepackage{epstopdf}
\usepackage{color}
\usepackage{tikz-cd}
\usepackage{bm}
\usepackage[colorlinks=true,linkcolor=blue,citecolor=red]{hyperref}%

\def\pa{\partial\Omega}
\def\P{{\mathbb P}}

\def\T{{\mathcal T}}

\def\erf{\mathrm{erf}}

\def\x{\bm{x}}

\begin{document}

\title{First-passage times to a fractal boundary: \\
local persistence exponent and its log-periodic oscillations}

\author{Yilin~Ye}
 \email{yilin.ye@polytechnique.edu}
\affiliation{
Laboratoire de Physique de la Mati\`{e}re Condens\'{e}e (UMR 7643), \\ 
CNRS -- Ecole Polytechnique, Institut Polytechnique de Paris, 91120 Palaiseau, France}

\author{Adrien~Chaigneau}
 \email{adrien.chaigneau@polytechnique.edu}
\affiliation{
Laboratoire de Physique de la Mati\`{e}re Condens\'{e}e (UMR 7643), \\ 
CNRS -- Ecole Polytechnique, Institut Polytechnique de Paris, 91120 Palaiseau, France}

\author{Denis~S.~Grebenkov}
 \email{denis.grebenkov@polytechnique.edu}
\affiliation{
Laboratoire de Physique de la Mati\`{e}re Condens\'{e}e (UMR 7643), \\ 
CNRS -- Ecole Polytechnique, Institut Polytechnique de Paris, 91120 Palaiseau, France}
\affiliation{CNRS - Universit\'e de Montr\'eal CRM - CNRS,
6128 succ Centre-Ville, Montr\'eal QC H3C 3J7, Canada}

\date{\today}

\begin{abstract}
We investigate the statistics of the first-passage time (FPT) to a
fractal self-similar boundary of the Koch snowflake.  When the
starting position is fixed near the absorbing boundary, the FPT
distribution exhibits an apparent power-law decay over a broad range
of timescales, culminated by an exponential cut-off.  By extensive
Monte Carlo simulations, we compute the local persistence exponent of
the survival probability and reveal its log-periodic oscillations in
time due to self-similarity of the boundary.  The effect of the
starting point onto this behavior is analyzed in depth.  Theoretical
bounds on the survival probability are derived from the analysis of
diffusion in a circular sector.  Physical rationales for the refined
structure of the survival probability are presented.
\end{abstract}




\keywords{diffusion, first-passage time, fractal, self-similarity, persistence exponent, survival probability}

\maketitle

\section{Introduction}
\label{sec:intro}

The statistics of first-passage times (FPTs) to various targets were
thoroughly investigated
\cite{Redner,Metzler,Masoliver,Oshanin,Dagdug}.  Most former works
were dedicated to a common setting when the starting point of the
particle is uniformly distributed, in which case the distribution of
the FPT is often close to be exponential \cite{Meyer11} and thus
determined by the mean FPT (MFPT).  The latter was studied in depth,
especially in the narrow-escape limit when the target is small
\cite{Singer06,Condamin07a,Benichou08,Agliari08,Pillay10,Mattos12,Benichou14,Holcman14,Benichou15,Grebenkov16,Simpson21}.
In this paper, we are interested in a different situation when the
particle starts in a vicinity of an absorbing self-similar boundary.
Here, the whole boundary plays the role of a sink and thus cannot be
treated as a small target so that most formerly used techniques and
results are not applicable anymore.

In the simplest case, one can think of diffusion in the upper
half-plane towards the absorbing horizontal axis.  As lateral
displacements are irrelevant due to the translational invariance of
the system in this direction, the problem is reduced to fairly
classical diffusion in the transverse direction on the positive
half-line $(0,\infty)$ with an absorbing endpoint at $0$, for which
the survival probability and the probability density function (PDF) of
the FPT are explicitly known:
\begin{subequations}  \label{eq:1D}
\begin{align}  \label{eq:St_1d}
S_{\rm 1d}(t) & = \erf\bigl(h_0/\sqrt{4Dt}\bigr), \\    \label{eq:Ht_1d}
H_{\rm 1d}(t) & = \frac{h_0}{\sqrt{4\pi Dt^3}}\, e^{-h_0^2/(4Dt)} \,,
\end{align}
\end{subequations}
where $\erf(z)$ is the error function, $h_0$ is the distance to the
absorbing axis, and $D$ is the diffusion coefficient.  In particular,
the survival probability exhibits a power-law decay at long times,
$S_{\rm 1d}(t) \propto t^{-\alpha}$, with the persistence exponent
$\alpha = 1/2$.  In disordered systems, there is no explicit solution
like Eq. (\ref{eq:St_1d}) but a power-law decay of the survival
probability is generally valid (see \cite{Bray13} and references
therein).  In particular, when the boundary is fractal, the
persistence exponent is affected by the geometric irregularity
\cite{Levitz06} (see also \cite{Rozanova12}).  For instance, in the
planar case, the long-time behavior is
\begin{equation}  \label{eq:S_fractal}
S(t) \propto t^{-\alpha},  \qquad  \alpha = d_f/2,
\end{equation}
where $d_f$ is the fractal (Minkowski) dimension of the boundary.
This result was rigorously obtained when the particle starts uniformly
on a contour line at a fixed small distance $h_0$ from the boundary.

\begin{figure}
\begin{center}
\includegraphics[width=88mm]{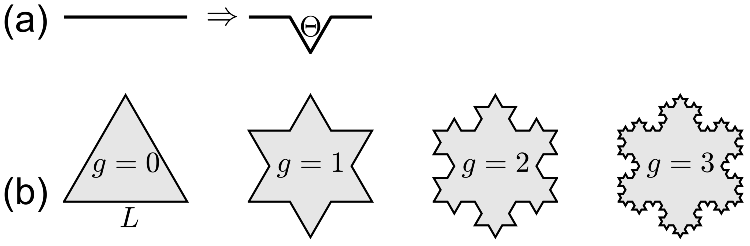} 
\end{center}
\caption{
Iterative construction of the Koch snowflake.  {\bf (a)} At each
iteration, each linear segment is replaced by a simple generator
formed by four equal segments and characterized by the angle $\Theta =
\pi/3$.  {\bf (b)} Starting from the zeroth generation $g = 0$ (an
equilateral triangle with the edges of length $L$), one constructs
iteratively next generations. }
\label{fig:koch_snowflakes}
\end{figure}

Our goal is to refine the above picture by considering a fixed
starting point near the boundary.  For this purpose, we study the
statistics of the FPT towards a fractal self-similar boundary of the
Koch snowflake (Fig. \ref{fig:koch_snowflakes}).  Former studies of
diffusion towards fractal boundaries were mainly focused on the
spatial distribution of the absorption points (so-called harmonic
measure or hitting probabilities)
\cite{Cates87,Mandelbrot90,Evertsz91,Evertsz92,Grebenkov05a,Grebenkov05b,Adams09},
that determines, e.g., the growth kinetics of diffusion-limited
aggregates \cite{Witten81,Halsey94,Grebenkov17w}.  We emphasize that
our setting is drastically different from diffusion in a disordered
fractal medium such as a percolation cluster \cite{Stauffer} or a
Sierpinski gasket.  In the latter case, the fractality of the domain
makes the diffusive dynamics anomalous that also impacts the
statistics of the FPT
\cite{OShaughnessy85,VandenBroech89,Condamin07,Condamin08,Haynes08,Benichou10,Lin10,Meroz11,Meyer12,Tavani16,Plyukhin16,Peng19,Levernier19,Hyman19,Zunke22,Baravi23}.  
In contrast, we consider ordinary diffusion inside a Euclidean
(non-fractal) domain (of dimension $2$) with a self-similar boundary
(of fractal dimension $1 < d_f < 2$).  Nevertheless, the behavior of
the survival probability and of the MFPT turns out to be much richer
than earlier expected.  For instance, the (average) persistence
exponent depends on the starting point and its neighborhood.
Moreover, we reveal a finer structure of the survival probability by
evaluating the local persistence exponent.  The latter is shown to
exhibit log-periodic oscillations in time.  In this light, the former
universal result (\ref{eq:S_fractal}) turns out to represent an
averaged behavior due to the uniform distribution of the starting
point.

The paper is organized as follows.  In Sec. \ref{sec:sector}, we
present analytical results on the FPT distribution in a sector and a
wedge.  These results will guide us in the analysis of the FPT to the
boundary of the Koch snowflake in Sec. \ref{sec:Koch}.  Section
\ref{sec:conclusion} concludes the paper.

\section{First-passage times in a sector}
\label{sec:sector}

It is instructive to start the analysis of the FPT distribution for
the zeroth generation of the Koch snowflake, namely, the equilateral
triangle (Fig. \ref{fig:koch_snowflakes}).  For this domain, the
Laplacian eigenfunctions with Dirichlet boundary condition are known
explicitly \cite{Lame,Pinsky80,Pinsky85,Grebenkov13}, which allows one
to write spectral expansions for the survival probability and the PDF.
However, we prefer to focus on a circular sector of angle $\Theta$ and
radius $R$.  When $\Theta = \pi/3$, this shape is geometrically close
to the equilateral triangle so that the FPT distribution exhibits very
similar features.  In turn, the circular sector gives more flexibility
(changing the angle $\Theta$) and more explicit formulas that
facilitate their interpretations.

\subsection{Circular sector}

In polar coordinates $(r,\theta)$, the sector is defined as
\begin{equation}
\Omega_{\Theta,R} = \{ (r,\theta)~:~ 0 < r < R, ~ 0 < \theta < \Theta\}.  
\end{equation}
The eigenvalues and eigenfunctions of the Laplace operator in
$\Omega_{\Theta,R}$ with Dirichlet boundary condition are given by
(see \cite{Grebenkov13} for details and references):
\begin{equation} 
\lambda_{nk} = \alpha_{nk}^2/R^2, \qquad
u_{nk} = c_{nk} J_{\nu_n}(\alpha_{nk} r/R) \sin(\nu_n \theta),
\end{equation}
where $\nu_n = \pi n/\Theta$ ($n = 1,2,3,\ldots$), $\alpha_{nk}$ are
the zeros of the Bessel function of the first kind,
$J_{\nu_n}(\alpha_{nk}) = 0$ (enumerated by $k = 1,2,3,\ldots$), and
$c_{nk}$ are the normalization coefficients:
\begin{align}  \nonumber
c_{nk}^{-2} & = \int\limits_0^R dr \, r \int\limits_0^{\Theta} d\theta \bigl[J_{\nu_n}(\alpha_{nk} r/R) \sin(\nu_n \theta)\bigr]^2 \\ 
& = \frac{R^2 \Theta}{4} J_{\nu_n+1}^2(\alpha_{nk}) .
\end{align}
As a consequence, the survival probability reads
\begin{align}  \nonumber
S_{\Theta,R}(t|\x_0) & = \sum\limits_{n=1}^\infty \sum\limits_{k=1}^\infty  
e^{-Dt\lambda_{nk}}\, u_{nk}(\x_0) \int\limits_{\Omega} d\x \, u_{nk}(\x) \\  \nonumber
& = \sum\limits_{n=1}^\infty \sum\limits_{k=1}^\infty  
e^{-Dt\lambda_{nk}}\, c_{nk}^2 J_{\nu_n}(\alpha_{nk} r_0/R) \sin(\nu_n\theta_0) \\    \label{eq:St_sector}
& \times   \frac{1-(-1)^n}{\nu_n} R^2  \int\limits_0^1 dx \, x \, J_{\nu_n}(\alpha_{nk} x) ,
\end{align}
where $\x_0 = (r_0,\theta_0)$ is the starting point in polar
coordinates.  In turn, the PDF of the FPT is obtained by taking the
time derivative:
\begin{equation}  \label{eq:Ht_sector}
H_{\Theta,R}(t|\x_0) = -\partial_t S_{\Theta,R}(t|\x_0).
\end{equation}
If the starting point $\x_0$ is uniformly distributed in $\Omega$, one
gets
\begin{align}  \nonumber
\overline{S_{\Theta,R}(t)} & = \frac{2}{\Theta R^2} \int\limits_{\Omega} d\x_0 \, S_{\Theta,R}(t|\x_0)  \\  \nonumber
& = \frac{2R^2}{\Theta} \sum\limits_{n=1}^\infty \sum\limits_{k=1}^\infty  e^{-Dt\lambda_{nk}}\, c_{nk}^2  \\
&  \times   \biggl(\frac{1-(-1)^n}{\nu_n} \int\limits_0^1 dx \, x \, J_{\nu_n}(\alpha_{nk} x)\biggr)^2 .
\end{align}  

In turn, the MFPT can be found as
\begin{align}  \nonumber
T_{\Theta,R}(\x_0) & = \int\limits_0^\infty dt \, t \, H_{\Theta,R}(t|\x_0) = \int\limits_0^\infty dt \, S_{\Theta,R}(t|\x_0) \\ \nonumber
& = \frac{R^2}{D} \sum\limits_{n=1}^\infty \sum\limits_{k=1}^\infty  
\frac{c_{nk}^2}{\lambda_{nk}} J_{\nu_n}(\alpha_{nk} r_0/R) \sin(\nu_n\theta_0) \\   \label{eq:MFPT_sector_formal}
& \times   \frac{1-(-1)^n}{\nu_n}   \int\limits_0^1 dx \, x \, J_{\nu_n}(\alpha_{nk} x) .
\end{align}
One can further simplify this expression by using the summation
identities for series with Bessel functions \cite{Watson,Grebenkov21}.
In particular, the Kneser-Sommerfeld expansion gives
\begin{equation}
\sum\limits_{k=1}^\infty \frac{J_{\nu}(\alpha_{\nu,k} z) J_\nu(\alpha_{\nu,k} z_0)}{\alpha_{\nu,k}^2 J_{\nu+1}^2(\alpha_{\nu,k})}
= \frac{z^\nu}{4\nu} (z_0^{-\nu} - z_0^\nu) 
\end{equation}
for $0 \leq z \leq z_0 \leq 1$ (see, e.g., Eq. (D10) from Table 3 in
\cite{Grebenkov21}; note that the factor $\pi/2$ was missing in the
right-hand side of Eq. (D10)).  As a consequence, we have
\begin{widetext}
\begin{align} \nonumber
T_{\Theta,R} 
& = \frac{R^2}{D \Theta} \sum\limits_{n=1}^\infty \frac{1-(-1)^n}{\nu_n} \sin(\nu_n \theta_0)  \frac{1}{\nu_n} 
\biggl\{\int\limits_0^{z_0} dz \,z \, z^{\nu_n} (z_0^{-\nu_n} - z_0^{\nu_n})  
+ \int\limits_{z_0}^1 dz \,z \, z_0^{\nu_n} (z^{-\nu_n} - z^{\nu_n}) \biggr\} \\   \nonumber
& = \frac{R^2}{D \Theta} \sum\limits_{n=1}^\infty \frac{1-(-1)^n}{\nu_n^2} \sin(\nu_n \theta_0)   
\biggl\{ \frac{z_0^{\nu_n+2}}{\nu_n+2} (z_0^{-\nu_n} - z_0^{\nu_n})  
+ z_0^{\nu_n} \biggl(\frac{1-z_0^{2-\nu_n}}{2-\nu_n} - \frac{1-z_0^{2+\nu_n}}{2+\nu_n}\biggr) \biggr\} \\  \label{:eq:MFPT_sector0}
& = \frac{2R^2}{D \Theta} \sum\limits_{n=1}^\infty \frac{1-(-1)^n}{\nu_n} \sin(\nu_n \theta_0) \frac{z_0^2 - z_0^{\nu_n}}{\nu_n^2-4} \,,   
\end{align}
\end{widetext}
where $z_0 = r_0/R$.  One can rewrite it as
\begin{equation}  \label{eq:MFPT_sector}
T_{\Theta,R} = T_{\Theta,\infty} - \frac{2R^2}{D \Theta} 
\sum\limits_{n=1}^\infty \frac{1-(-1)^n}{\nu_n (\nu_n^2-4)} \sin(\nu_n \theta_0) (r_0/R)^{\nu_n} \,,
\end{equation}
where 
\begin{equation}  \label{eq:MFPT_wedge}
T_{\Theta,\infty} = \frac{r_0^2}{4D} \biggl(\frac{\cos (2\theta_0-\Theta)}{\cos\Theta} - 1\biggr).
\end{equation}
We could not find earlier references with an explicit form
(\ref{eq:MFPT_sector}) for the MFPT in a sector.  When $\Theta <
\pi/2$, the expression (\ref{eq:MFPT_wedge}) can be interpreted as the
MFPT to the wedge of angle $\Theta$ (or infinite sector with $R =
\infty$) that was discussed in \cite{LeVot20}.  In turn, if $\Theta
\geq \pi/2$, the MFPT to the wedge is infinite (see below), whereas
Eq. (\ref{eq:MFPT_wedge}) yields negative values.  At the same time,
Eq. (\ref{eq:MFPT_sector}) is valid for any $\Theta$ from $0$ to
$2\pi$.

This important result unveils the scaling of the MFPT with respect to
$r_0$.  As $r_0 \to 0$, the sum in Eq. (\ref{eq:MFPT_sector}) scales
in the leading order as $r_0^{\pi/\Theta}$, whereas
$T_{\Theta,\infty}$ scales as $r_0^2$.  Depending on $\Theta$, one of
these two contributions is dominant, implying
\begin{equation}  \label{eq:TTheta_asympt}
\frac{D}{R^2} T_{\Theta,R} \approx \begin{cases} \frac{1}{4} \bigl(\frac{\cos (2\theta_0-\Theta)}{\cos\Theta} - 1\bigr) 
(r_0/R)^2  \quad (\Theta < \pi/2), \cr
\frac{\sin(2\theta_0)}{\pi} (r_0/R)^2 \ln(R/r_0)  \hskip 6mm (\Theta = \pi/2), \cr
\frac{\Theta^2 \sin(\pi \theta_0/\Theta)}{\pi (\Theta^2 - \pi^2/4)}  \, (r_0/R)^{\pi/\Theta} \qquad (\Theta > \pi/2) \end{cases}
\end{equation}
(in the border case $\Theta = \pi/2$, both contributions are
comparable so that one has to use directly
Eq. (\ref{:eq:MFPT_sector0}) to get the logarithmic correction).  The
exact form of the second moment of the FPT and the discussion on its
variance are reported in Appendix \ref{sec:T2}.

Figure \ref{fig:MFPT_sector} shows the behavior of $T_{\Theta,R}$ as a
function of $r_0$ and illustrates the accuracy of the asymptotic
relations (\ref{eq:TTheta_asympt}).  As the MFPT vanishes in the
limits $r_0 = 0$ and $r_0 = R$, there is an optimal location of the
starting point that maximizes $T_{\Theta,R}$ with respect to $r_0$.
In the considered case when the starting point is on the middle ray
(i.e., $\theta_0 = \Theta/2$), one sees that the optimal value
$r_0^{*}$, at which the MFPT is maximal, decreases as $\Theta$
increases (i.e., the maximum of the MFPT shifts to the left).  Indeed,
the MFPT is expected to be larger when the starting point is further
from the boundary.  For the starting point on the middle ray, the
largest distance to the boundary is achieved from the center of the
incircle (i.e., the largest circle inscribed into the circular
sector).  Its radius $\rho$ can be found by setting $\rho = r_0^*
\sin(\Theta/2)$ and $r_0^* + \rho = R$, from which $\rho = R
\sin(\Theta/2)/(1 + \sin(\Theta/2))$ and thus $r_0^* =
R/(1+\sin(\Theta/2))$ for $\Theta \leq \pi$.  For instance, we get
$r_0^* \approx 0.79,~ 0.67,~ 0.54$ at $\Theta = \pi/6,~\pi/3,~
2\pi/3$, respectively, which are in excellent agreement with the
positions of the maxima shown in Fig. \ref{fig:MFPT_sector}.  For any
$\Theta > \pi$, the inradius $\rho$ remains equal to $1/2$, so that
the optimal position $r_0^*$ is also located around $0.5$.  Note that
these extremal properties are not captured by the asymptotic relations
(\ref{eq:TTheta_asympt}).

\begin{figure}
\begin{center}
\includegraphics[width=88mm]{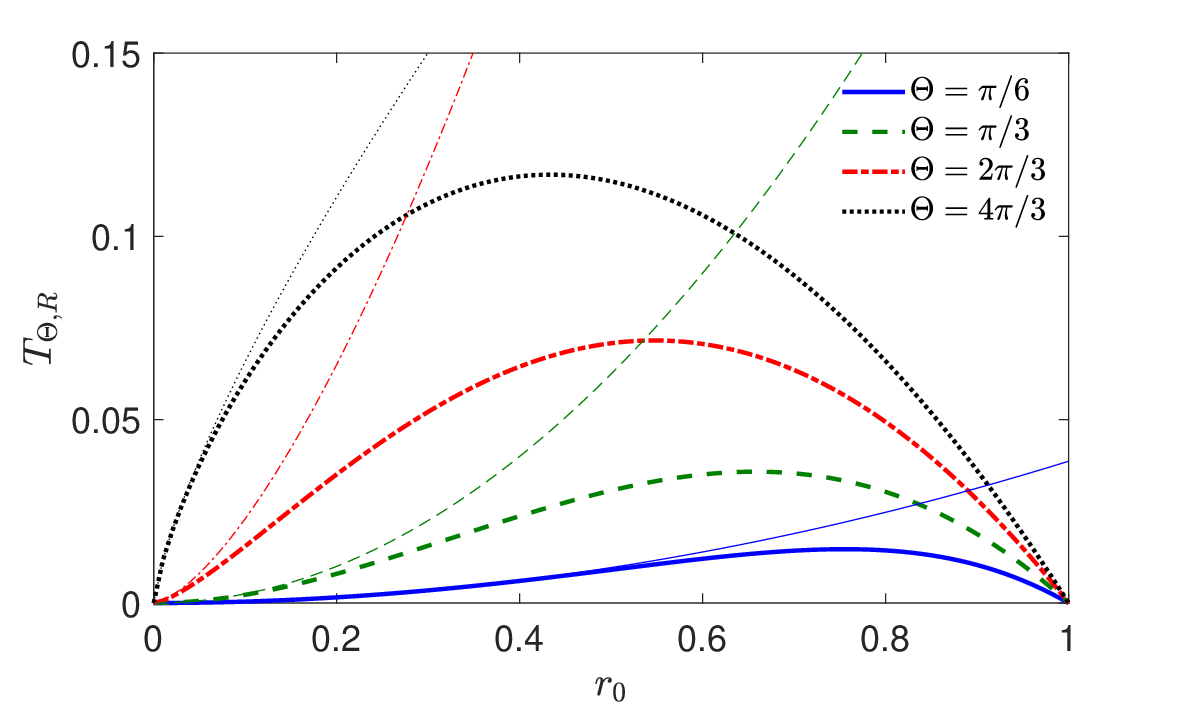} 
\end{center}
\caption{
MFPT $T_{\Theta,R}$ in sectors of different angles $\Theta$, with
$\theta_0 = \Theta/2$ and variable $r_0$.  Thin lines present the
asymptotic relations (\ref{eq:TTheta_asympt}).  We set $D = 1$ and $R
= 1$.}
\label{fig:MFPT_sector}
\end{figure}

\subsection{Wedge case}

For the wedge of angle $\Theta$ (i.e., the infinite sector with $R =
\infty$), one can get an explicit solution for the propagator and the
survival probability (see
\cite{Considine89,Comtet03,Chupeau15,LeVot20} and references therein):
\begin{equation}  \label{eq:St_wedge}
S_{\Theta,\infty}(t|\x_0) = 2 \sum\limits_{n=1}^\infty \frac{1-(-1)^n}{\pi n} \sin(\nu_n \theta_0) R_{\nu_n}\bigl(r_0/\sqrt{Dt}\bigr),
\end{equation}
where 
\begin{equation}
R_\nu(z) = \frac{\sqrt{\pi}}{4} z e^{-z^2/8} \biggl(I_{\frac{\nu-1}{2}}(z^2/8) + I_{\frac{\nu+1}{2}}(z^2/8)\biggr),
\end{equation}
with $I_\nu(z)$ being the modified Bessel function of the first kind
(when the wedge angle has a special form, $\Theta = \pi/m$, with an
integer $m$, the above expression can be further simplified, see
\cite{Dy08}, as well as \cite{Dy13} for an extension).  Note that
$R_\nu(z)\to 1$ as $z\to \infty$ so that $S_{\Theta,\infty}(t|\x_0)$
approaches $1$ as $t\to 0$, as it should.  In turn, in the long-time
limit, one retrieves the power-law decay \cite{Redner}
\begin{equation}  \label{eq:St_wedge_long}
S_{\Theta,\infty}(t|\x_0)\propto (r_0/\sqrt{Dt})^{\pi/\Theta}  \qquad (t \gg r_0^2/D), 
\end{equation}
with the persistence exponent $\alpha = \pi/(2\Theta)$.  As a
consequence, the MFPT is infinite for $\Theta \geq \pi/2$ and finite
for $\Theta < \pi/2$ (see Eq. (\ref{eq:MFPT_wedge})).   The
negative derivative of $S_{\Theta,\infty}(t|\x_0)$ with respect to $t$
yields the probability density function of the FPT to the wedge
boundary:
\begin{align}  \nonumber
H_{\Theta,\infty}(t|\x_0) & = \frac{r_0}{\sqrt{Dt^3}}
\sum\limits_{n=1}^\infty \frac{1-(-1)^n}{\pi n} \\    \label{eq:Ht_wedge}
& \times \sin(\nu_n \theta_0) R'_{\nu_n}\bigl(r_0/\sqrt{Dt}\bigr),
\end{align}
where 
\begin{equation}
R'_\nu(z) = \frac{\sqrt{\pi}}{4} \nu  e^{-z^2/8} \biggl(I_{\frac{\nu-1}{2}}(z^2/8) - I_{\frac{\nu+1}{2}}(z^2/8)\biggr)
\end{equation}
is the derivative of $R_\nu(z)$.

\subsection{Three distinct regimes}

Figure \ref{fig:St_sector} illustrates the behavior of the survival
probability and the PDF of the FPT for the sector of angle $\Theta =
\pi/3$ and radius $R = 1$.  We choose the starting point $\x_0 =
(r_0,\theta_0)$ to be located near the horizontal segment of the
sector in the vicinity of its vertex (with $r_0 = 10^{-2}$ and
$\theta_0 = 10^{-2}$ so that the initial distance to the boundary,
$h_0 = r_0 \sin\theta_0 \approx 10^{-4}$, is the smallest
lengthscale).  This choice helps to illustrate three distinct regimes:

\begin{figure}
\begin{center}
\includegraphics[width=88mm]{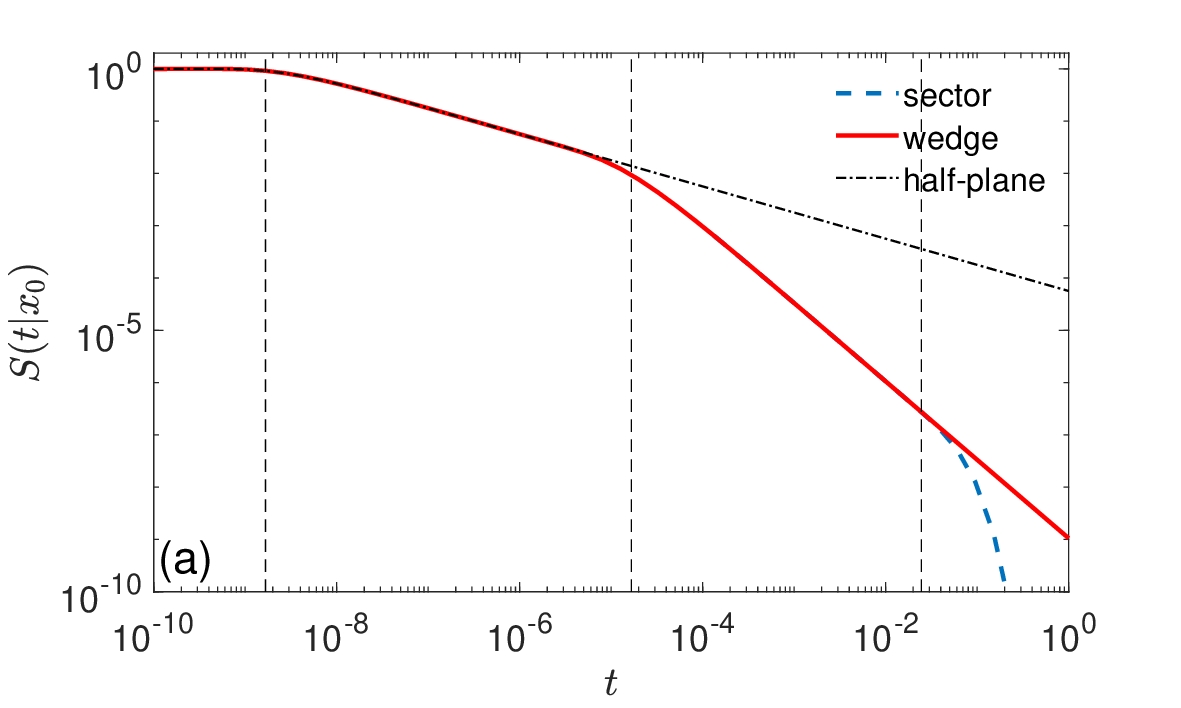} 
\includegraphics[width=88mm]{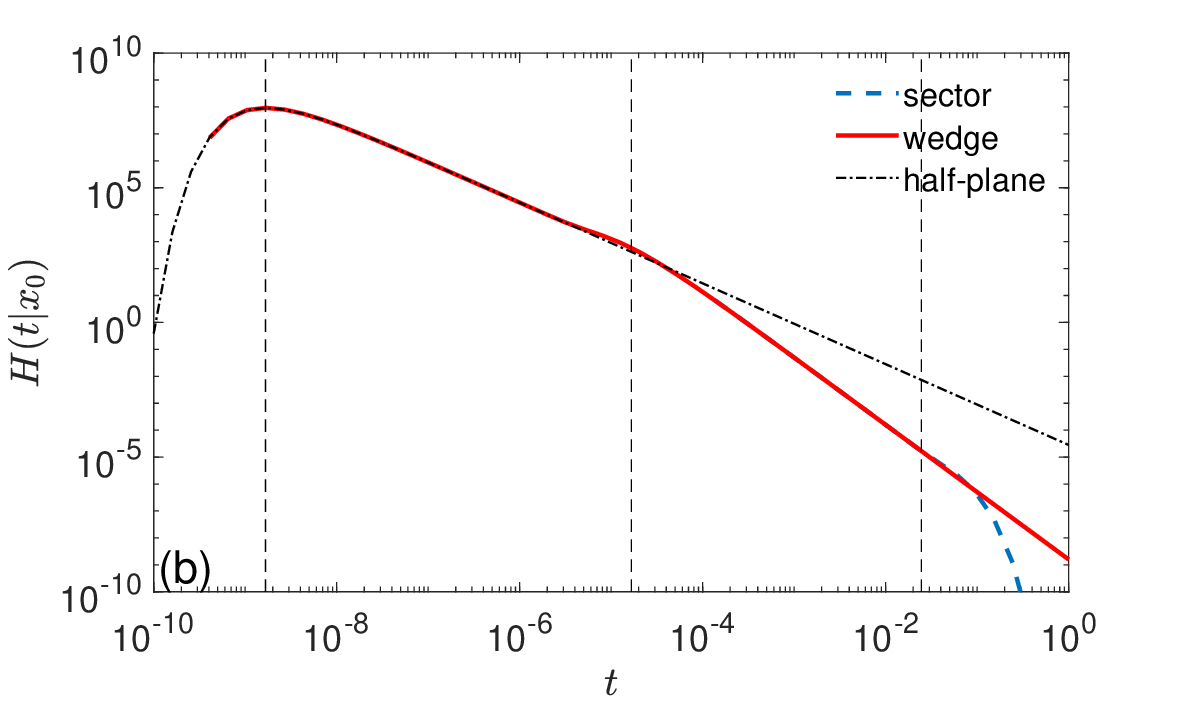} 
\end{center}
\caption{
{\bf (a)} Survival probability $S_{\Theta,R}(t|\x_0)$ in a circular
sector of angle $\Theta = \pi/3$ and radius $R = 1$, with the starting
point $\x_0 = (r_0,\theta_0) = (10^{-2}, 10^{-2})$ in polar
coordinates, and $D = 1$.  This survival probability, which is shown
by thick dashed line, was obtained by truncating the double series in
Eq. (\ref{eq:St_sector}) to 7354 terms to enable its accurate
computation even at very short times.  It is compared to
$S_{\Theta,\infty}(t|\x_0)$ for the wedge (thick solid line,
Eq. (\ref{eq:St_wedge})) and to $S_{\rm 1d}(t)$ in the half-plane
(thin dash-dotted line, Eq. (\ref{eq:St_1d})).  {\bf (b)} PDF
$H_{\Theta,R}(t|\x_0)$ in the same sector (shown by thick dashed line)
was also computed from its truncated spectral expansion; it is
compared to $H_{\Theta,\infty}(t|\x_0)$ (thick solid line,
Eq. (\ref{eq:Ht_wedge})) and $H_{\rm 1d}(t)$ (dash-dotted line,
Eq. (\ref{eq:Ht_1d})).  Three vertical dashed lines indicate three
timescales: $h_0^2/(6D)$, $r_0^2/(6D)$, and $1/(D\lambda_{\rm min})
\approx 0.025\, R^2/D$, where $h_0 = r_0 \sin \theta_0 \approx
10^{-4}$.  }
\label{fig:St_sector}
\end{figure}

(i) At short times, $t \ll r_0^2/D$, the particle diffuses near the
absorbing horizontal segment, as if it was in the upper half-plane
above the absorbing horizontal line.  In this regime, the survival
probability and the PDF are determined by diffusion in the transverse
direction on the semi-axis $(0,+\infty)$ with the absorbing endpoint,
so that $S_{\Theta,R}(t|\x_0) \approx S_{\rm 1d}(t)$ and
$H_{\Theta,R}(t|\x_0) \approx H_{\rm 1d}(t)$, given by
Eqs. (\ref{eq:1D}).  This short-time behavior, which is shown by a
dash-dotted line, is universal (see \cite{Godec16,Grebenkov18d} for
more discussions).  In particular, if $h_0^2/D \ll t \ll r_0^2/D$, one
has $S_{\Theta,R}(t|\x_0) \approx h_0/\sqrt{\pi Dt}$.  Note also that
the maximum of the L\'evy-Smirnov density in the right-hand side of
Eq. (\ref{eq:Ht_1d}) is realized at $t_{\rm mp} = h_0^2/(6D)$.  In
this setting, one sees that the most probable FPT in the sector is
well approximated by that $t_{\rm mp}$.

(ii) At intermediate times, $r_0^2/D \lesssim t \ll R^2/D$, the
particle is confined within an absorbing angle that affects its
survival.  In turn, it does not ``feel'' the presence of the arc of
the sector (located at distance $R$), as if it diffused in an infinite
wedge.  As a consequence, the survival probability and the PDF of the
FPT for the sector are very close to that for the wedge (shown by
solid lines).  Figure \ref{fig:St_sector} confirms the remarkable
agreement between these quantities over intermediate times.  In
particular, one observes the power-law asymptotic decay
(\ref{eq:St_wedge_long}) at $r_0^2/D \ll t \ll R^2/D$.

(iii) At long times, $t \gtrsim R^2/D$, the particle explores the
whole sector and thus starts to ``feel'' the confinement by the
absorbing arc.  The confinement inside a bounded domain changes the
previous power-law regime, $S_{\Theta,R}(t|\x_0) \propto
t^{-\pi/(2\Theta)}$, into an exponential cut-off, which is determined
by the smallest eigenvalue $\lambda_{\rm min}$ of the Laplace
operator: $\lambda_{\rm min} = \alpha_{1,1}^2/R^2$, where
$\alpha_{1,1}$ is the smallest positive zero of the Bessel function
$J_{\pi/\Theta}(z)$.  In fact, when $t \gg R^2/D$, one has
$S_{\Theta,R}(t|\x_0) \propto e^{-Dt\lambda_{\rm min}}$.  For $\Theta
= \pi/3$, one gets $\alpha_{1,1} \approx 6.3802$ that determines the
decay rate $1/(D\lambda_{\rm min}) \approx 0.025 R^2/D$.

The three relevant timescales revealed in the above analysis are
naturally related to three lengthscales: the distance $h_0$ to the
flat boundary, the distance $r_0$ to a singularity (the angle), and
the radius $R$ of the sector.  As a consequence, the timespans of the
three regimes depend on the location of the starting point.  For
instance, changing $\theta_0$ from $10^{-2}$ to, say, $\pi/6$ (the
middle ray of the sector) can eliminate the first regime (as $h_0$
becomes comparable to $r_0$), whereas changing $r_0$ from $10^{-2}$
to, say, $0.5$ can eliminate the second regime (as $r_0$ becomes
comparable to $R$).  Finally, if the starting point is located
somewhere in the middle of the sector, both first and second regimes
can be effectively removed.  In the same vein, if the starting point
is uniformly distributed in the sector, the first two regimes become
irrelevant.  In this uniform setting, the MFPT is close to the decay
rate that determines the exponential cut-off, whereas the FPT obeys
approximately an exponential law \cite{Meyer11}.  Most former works
dealt with this setting and focused on the MFPT as the unique relevant
timescale.  The aim of the present paper is to investigate the
opposite situation when the particle starts in a vicinity of the
absorbing boundary, in which case the FPT distribution is much richer.

\section{First-passage times in the Koch snowflake}
\label{sec:Koch}

Despite the geometric simplicity of the sector domain, the three
regimes discussed in Sec. \ref{sec:sector} capture the generic
features of the FPT distribution.  When the sector is replaced by a
polygon, the first and the third regimes remain unchanged.  Indeed, if
the particle starts very close to the boundary, one still deals with
diffusion near a segment at very short times, so that
Eqs. (\ref{eq:1D}) provide accurate approximations for $S(t|\x_0)$ and
$H(t|\x_0)$.  Moreover, the exponential cut-off is also present at
very long times due to confinement in a bounded domain.  Both limits
are rather basic and well understood.  For instance, several studies
focused on the smallest eigenvalue $\lambda_{\rm min}$ of the Laplace
operator in domains with fractal boundary (see
\cite{Sapoval91,Lapidus95,Even99,Daudert07,Banjai07} and references
therein) that determines the decay time of the exponential cut-off
(see below).  Our main emphasis is therefore on the intermediate
regime.  For this purpose, we choose the starting point to be at
distance $h_0$ from the boundary such that
\begin{equation}  \label{eq:cond}
\ell_g \ll h_0 \ll L, 
\end{equation}
where $L$ is the size of the domain, and $\ell_g$ is the length of the
smallest segment used for constructing the prefractal boundary of
generation $g$, as explained below.

\subsection{Koch snowflake}

To enable efficient numerical simulations, we consider diffusion in
the Koch snowflake that is constructed iteratively from a simple
generator with a chosen angle $\Theta$
(Fig. \ref{fig:koch_snowflakes}a).  One starts from an equilateral
triangle $\Omega_0$ with edges of length $L$ that is referred to as
the generation $0$.  Each linear segment of the boundary $\pa_0$ is
replaced by rescaled and appropriately rotated generator to produce
the first generation $\Omega_1$.  Its boundary $\pa_1$ is composed of
$3\cdot 4 = 12$ segments of length $\ell_1 = L/(2[1 +
\sin(\Theta/2)])$.  In the same way, one constructs iteratively the
successive generations $\Omega_2, \Omega_3, \ldots$ of the Koch
snowflake (Fig. \ref{fig:koch_snowflakes}b).  The boundary $\pa_g$ is
composed of $3\cdot 4^g$ segments of length
\begin{equation}
\ell_g = \frac{L}{(2[1 + \sin(\Theta/2)])^g} \,.
\end{equation}
The boundary $\pa_\infty$ of the limiting domain $\Omega_\infty$ is
fractal; in particular, it is not differentiable and has infinite
length (in fact, the perimeter of $\pa_g$, $3\cdot 4^g \ell_g$,
diverges as $g\to \infty$).  By a scaling argument, the fractal
dimension of $\pa_\infty$ is
\begin{equation}  \label{eq:df}
d_f = \frac{\ln(4)}{\ln(2[1+\sin(\Theta/2)])} \,, 
\end{equation}
and it ranges from $1$ at $\Theta = \pi$ to $2$ as $\Theta \to 0$.
This family of Koch snowflakes is therefore convenient for studying
the effect of a self-similar absorbing boundary onto the FPT
distribution.  Throughout this paper, we focus on the canonical Koch
snowflake with $\Theta = \pi/3$ so that $\ell_g = L/3^g$ and $d_f =
\ln(4)/\ln(3) \approx 1.262$.

We study the distribution of the FPT, $\T_g$, to the boundary $\pa_g$
of the $g$-th generation $\Omega_g$.  The associated survival
probability and the PDF are denoted as $S_g(t|\x_0) =
\P_{\x_0} \{\T_g > t\}$ and $H_g(t|\x_0) = - \partial_t S_g(t|\x_0)$,
respectively.  Note that these quantities also determine all the
moments of the FPT, which are finite for any bounded domain
$\Omega_g$; for instance, the MFPT is
\begin{equation}  \label{eq:MFPT_def}
\langle \T_g \rangle_{\x_0} = \int\limits_0^\infty dt \, t \, H_g(t|\x_0) = \int\limits_0^\infty dt \, S_g(t|\x_0).
\end{equation}

\subsection{Description of Monte Carlo simulations}
\label{sec:numerics}

In order to estimate these quantities, we employ the geometry-adapted
fast random walk algorithm \cite{Grebenkov05a,Grebenkov05b}, which
combines the standard walk-on-spheres method by Muller \cite{Muller56}
and self-similar structure of the Koch snowflake for a rapid
evaluation of the distance to the boundary.  For a given $g$, a random
trajectory starts from a fixed point $\x_0$ and continues until the
distance to the boundary $\pa_g$ becomes smaller than a prescribed
(very small) threshold $\epsilon$.  The duration of such a trajectory
approximates the FPT to the absorbing boundary $\pa_g$.  Repeating
this simulation $N$ times, one gets an empirical statistics of the FPT
that allows one to estimate both $S_g(t|\x_0)$ and $H_g(t|\x_0)$.  For
all simulations, we used $\epsilon = 10^{-10}$ to ensure very accurate
approximation of the FPT even for large $g$, and $N = 10^8$ to get a
sufficient statistics for estimating $S_g(t|\x_0)$ and $H_g(t|\x_0)$.
In order to get more accurate statistics of rare realizations of large
FPTs, we also performed additional simulations, in which the generated
FPT $\T_g$ below a prescribed time $T_{\rm min}$ were discarded to
accumulate $N = 10^8$ realizations of the FPT above $T_{\rm min}$.  A
home-built code was written in Fortran90.  Numerous datasets of
simulated FPTs were recorded for different generations $g$ and
starting points $\x_0$.  Each dataset was then analyzed in Matlab by
using the standard routines \verb|ecdf| and \verb|hist| for getting
the empirical cumulative distribution function (CDF) and the histogram
(the latter was then rescaled into an empirical PDF).  As the
distribution of the FPT $\T_g$ is very broad, we actually produced the
histogram of $\ln (\T_g)$ and then transformed it into the histogram
of $\T_g$.  Throughout this section, we fix length and time units by
setting $L = 2$ and $D = 1$.

For the zeroth generation $\Omega_0$ (i.e., the equilateral triangle
of lengthside $L = 2$), the smallest eigenvalue of the Dirichlet
Laplacian is known exactly: $\lambda_{\rm min} = 4\pi^2/3 \approx
13.16$, see \cite{Pinsky80}.  As $g$ increases, $\lambda_{\rm min}$
decreases and rapidly converges to a well-defined limit.  Banjai
computed numerically the smallest eigenvalue of the Laplace operator
for various generations and showed that it changes very little for $g
\geq 4$ (see Table 1 in \cite{Banjai07}).  His numerical result yields
$\lambda_{\rm min} \simeq 9.837$ for our setting with $L = 2$.  This
value determines the confinement timescale:
\begin{equation}  \label{eq:t2}
t_c = \frac{1}{D\lambda_{\rm min}} \approx 0.1.
\end{equation}

\subsection{MFPT}

\begin{figure}
\includegraphics[width=88mm]{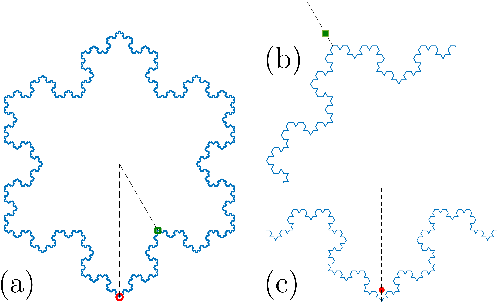}
\caption{
{\bf (a)} The 10th generation of the Koch snowflake.  Circle and
square indicate two starting points close to two vertices of angles
$\pi/3$ and $4\pi/3$, respectively.  {\bf (b,c)} Two panels on the
right show a zoom around each of these starting points.}
\label{fig:koch_g10}
\end{figure}

\begin{figure}
\includegraphics[width=88mm]{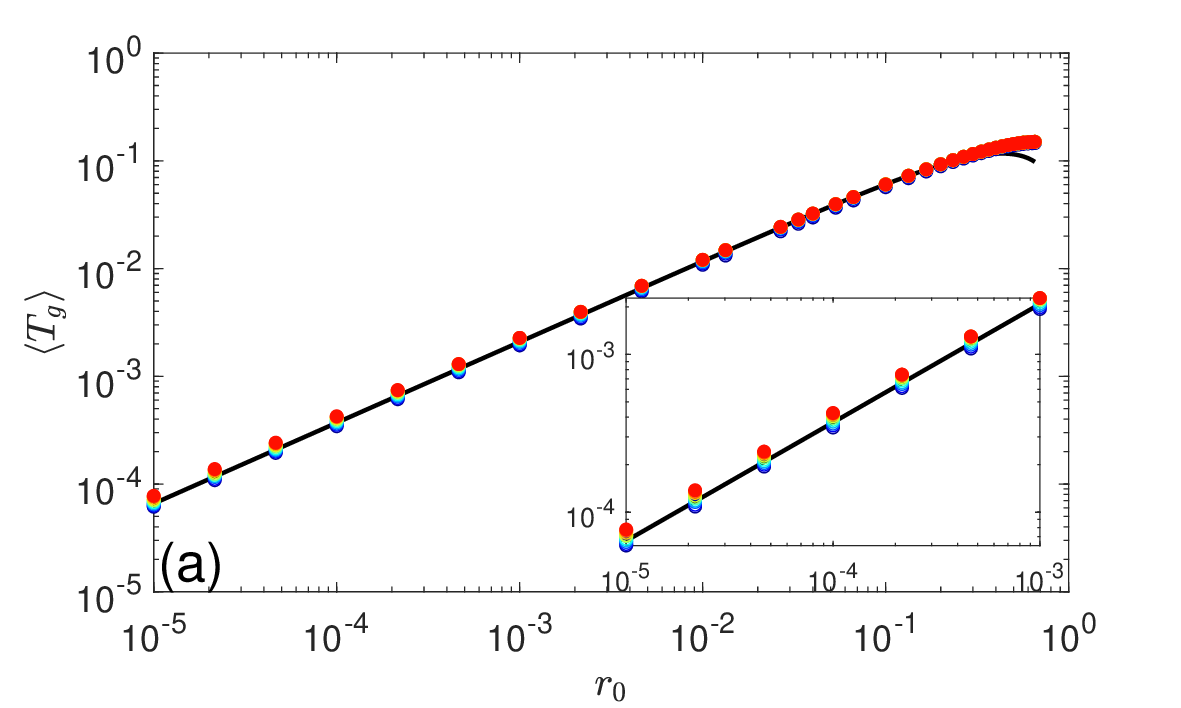} 
\includegraphics[width=88mm]{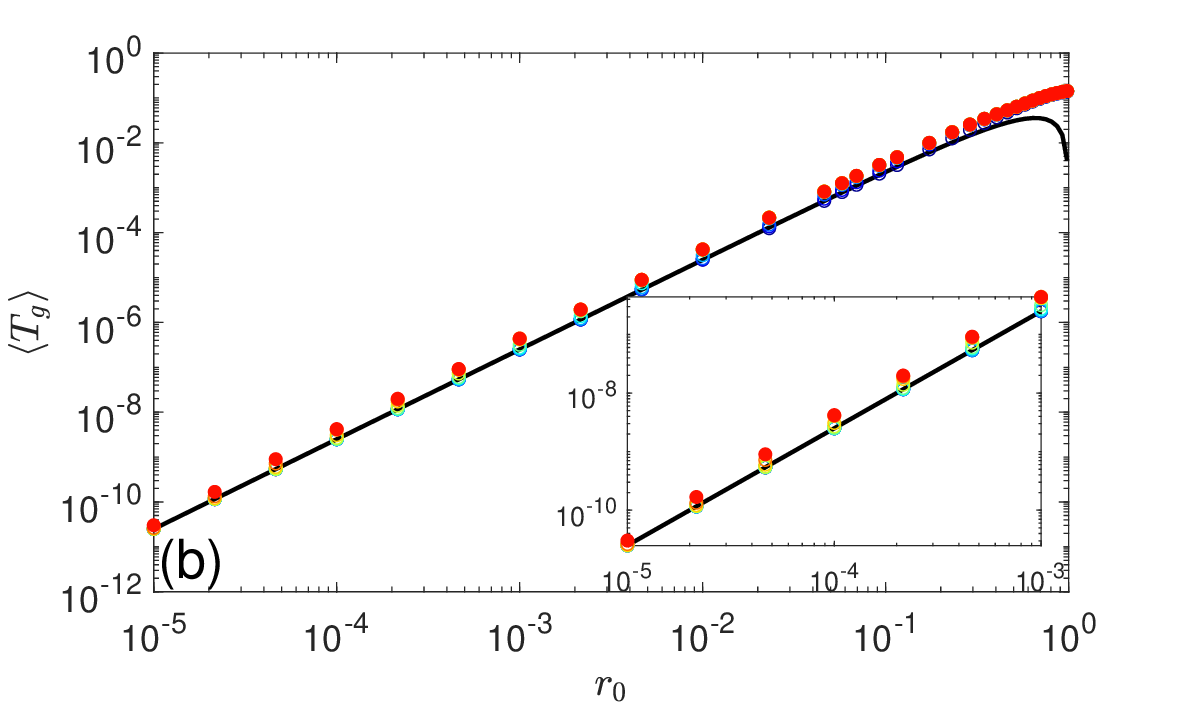} 
\caption{
MFPT as a function of the distance $r_0$ to a chosen vertex of the
boundary, for different generation of the Koch snowflake.  {\bf (a)}
Approaching a vertex of angle $4\pi/3$, see
Fig. \ref{fig:koch_g10}(b); {\bf (b)} approaching a vertex of angle
$\pi/3$, see Fig. \ref{fig:koch_g10}(c).  Solid line presents the MFPT
$T_{\Theta,R}(t|\x_0)$ in the sector of radius $R = 1$ that exhibits
the small-$r_0$ scaling behaviors $r_0^{3/4}$ (panel {\bf (a)}) and
$r_0^2$ (panel {\bf (b)}).  Color of symbols changes from dark blue
($g = 1$) to red ($g = 10$), the last one being filled.  Insets zoom
the regions of small $r_0$.}
\label{fig:MFPT}
%
\end{figure}

We start by evaluating the MFPT as a function of the distance $r_0$ to
a chosen vertex of the boundary.  For this purpose, we choose a ray in
the middle of the angle and change the distance $r_0$ to the vertex,
as illustrated on Fig. \ref{fig:koch_g10}(b,c).  Panels (a) and (b) of
Fig. \ref{fig:MFPT} show the dependence of the MFPT on $r_0$ for two
vertices of angles $4\pi/3$ and $\pi/3$, respectively.  In both cases,
one sees that the MFPT is close to $T_{\Theta,R}(t|\x_0)$ in the
sector and thus inherits its power-law scaling, $\langle
\T_g\rangle_{\x_0} \propto r_0^\beta$, with $\beta = 3/4$ and $\beta =
2$ for two considered cases, respectively.  The dependence of the MFPT
on the generation $g$ is rather weak.

\subsection{Theoretical bounds}

Let us now switch to the analysis of the survival probability and
first discuss a simple way to get its theoretical bounds.  Let
$S(t|\x_0)$ and $S'(t|\x_0)$ denote the survival probabilities in two
domains $\Omega$ and $\Omega'$ such that $\x_0 \in \Omega' \subset
\Omega$.  Any random trajectory that started from $\x_0$ and has
reached the boundary $\pa$ must hit or cross the boundary $\pa'$ of
the smaller enclosed domain $\Omega'$.  As a consequence, the FPT
$\tau'$ to $\pa'$ is {\it shorter} than the FPT $\tau$ to $\pa$.  More
rigorously, the probabilistic event $\{\tau' > t\}$ is included into
the probabilistic event $\{ \tau > t\}$, and thus
\begin{equation}  \label{eq:S_bound}
S'(t|\x_0) \leq S(t|\x_0)  \qquad \textrm{for any~} t\geq 0.  
\end{equation}
The integral of this inequality over $t$ yields the intuitively
expected inequality $\langle \tau'\rangle \leq \langle \tau \rangle$,
i.e., the MFPT $\langle \tau'\rangle$ to the smaller enclosed boundary
$\pa'$ is smaller than the MFPT $\langle \tau\rangle$ to the larger
boundary $\pa$.

For instance, if $\Omega'$ and $\Omega$ are two sectors of the same
angle $\Theta$ and radii $R' < R$, the survival probability in the
shorter sector is smaller: $S_{\Theta,R'}(t|\x_0) \leq
S_{\Theta,R}(t|\x_0)$.  Moreover, if $\Omega$ is the wedge of the
angle $\Theta$, then the survival probability
$S_{\Theta,\infty}(t|\x_0)$ in the wedge, given by
Eq. (\ref{eq:St_wedge}), is the upper bound for the survival
probability $S_{\Theta,R}(t|\x_0)$ in the sector (and also for the
equilateral triangle).  Accordingly, the MFPT in the wedge, given by
Eq. (\ref{eq:MFPT_wedge}), is the upper bound for the MFPT in the
sector, given by Eq. (\ref{eq:MFPT_sector}).
In the same vein, since $\Omega_{g-1} \subset \Omega_g$, one also has
\begin{equation}
S_{g-1}(t|\x_0) \leq S_g(t|\x_0)  \quad \Rightarrow  \quad \langle \T_{g-1}\rangle_{\x_0} \leq \langle \T_g\rangle_{\x_0}.
\end{equation}

\begin{figure}
\begin{center}
\includegraphics[width=40mm]{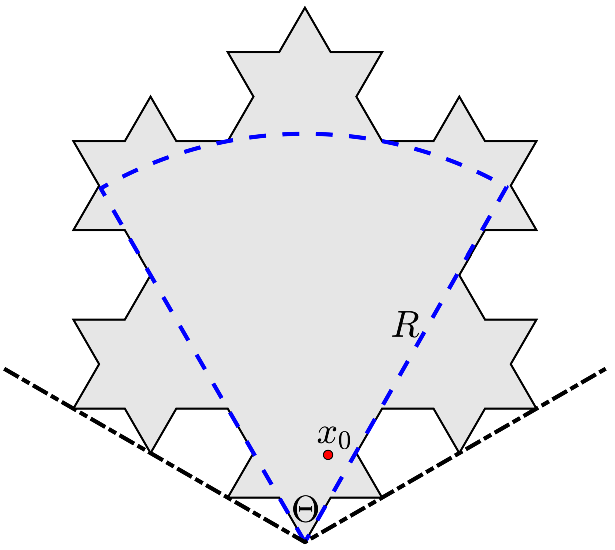} 
\end{center}
\caption{
The second generation of the Koch snowflake of angle $\Theta =\pi/3$
and $L = 2$, with the largest inscribed sector of radius $R =
L\sqrt{7}/3$ (blue dashed line) and the larger wedge of angle $\Theta'
= 2\pi/3$ (black dash-dotted line).  Red circle indicates the starting
point $\x_0$.}
\label{fig:bounds}
\end{figure}

Similarly, one can inscribe a circular sector of angle $\Theta$ and
radius $R$ into $\Omega_g$ to get a {\it lower bound} for
$S_g(t|\x_0)$:
\begin{equation}
S_{\Theta,R}(t|\x_0) \leq S_g(t|\x_0)  \qquad \textrm{for any~} t\geq 0,
\end{equation}
where the largest possible radius is $R = L\sqrt{7}/3$ for $\Theta =
\pi/3$ (Fig. \ref{fig:bounds}).  Moreover, if $\x_0$ lies close to the
vertex of angle $\Theta$, i.e., $r_0 \ll R$, the survival probability
$S_{\Theta,R}(t|\x_0)$ can be accurately approximated by
$S_{\Theta,\infty}(t|\x_0)$ for the wedge, at least for intermediate
times $t \ll L^2/D$.

At the same time, the Koch snowflake can be inscribed into a larger
wedge of angle $\Theta' = (\pi + \Theta)/2$ (Fig. \ref{fig:bounds}).
As a consequence, the survival probability
$S_{\Theta',\infty}(t|\x_0)$ for the larger wedge is an upper bound
for $S_g(t|\x_0)$:
\begin{equation}
S_g(t|\x_0) \leq S_{\Theta',\infty}(t|\x_0).
\end{equation}

\subsection{Survival probability and PDF}

\begin{figure}
\begin{center}
\includegraphics[width=88mm]{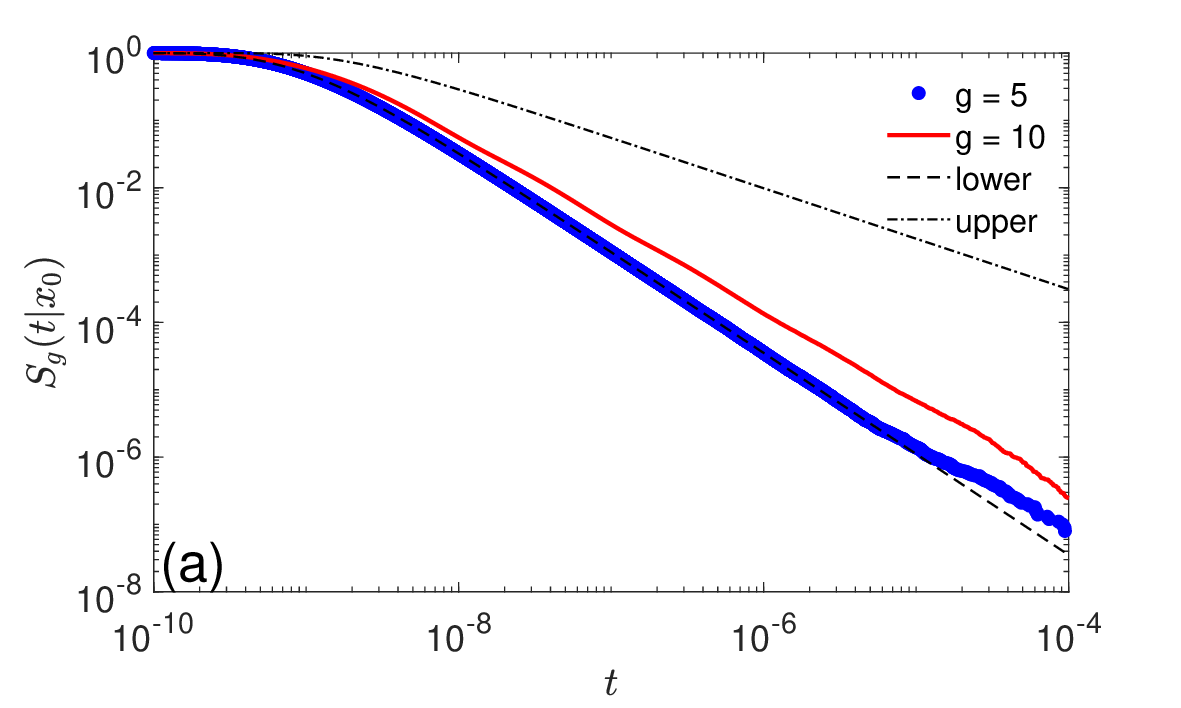} 
\includegraphics[width=88mm]{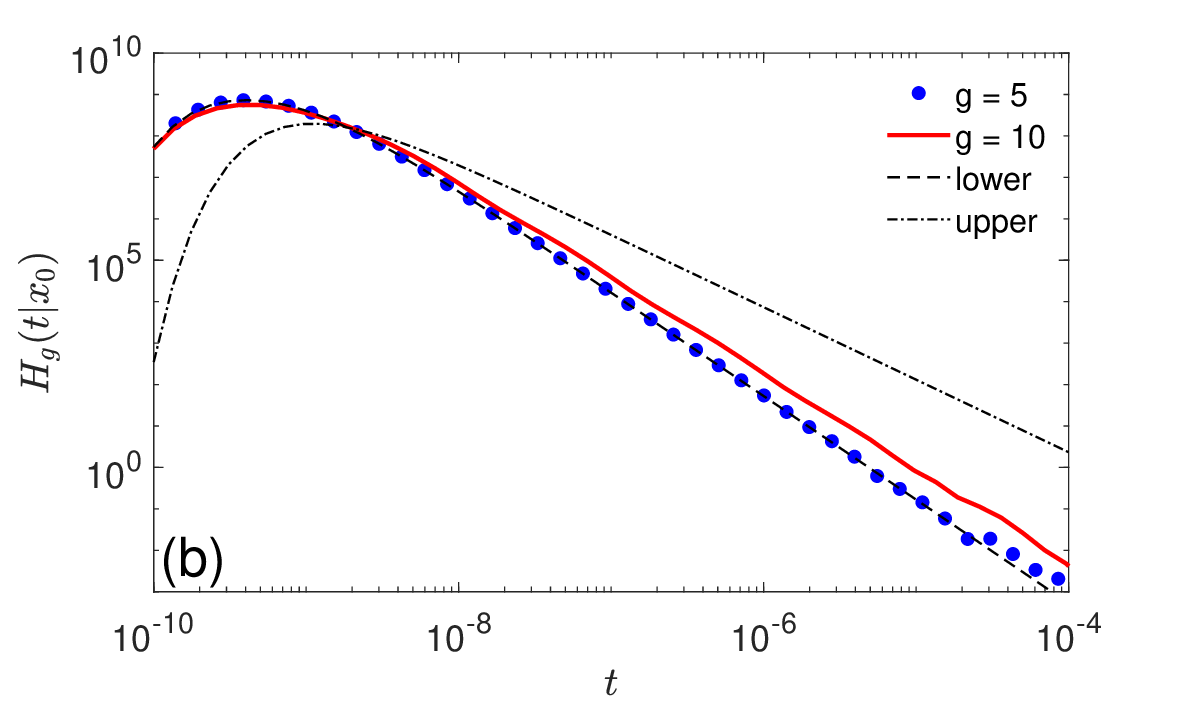} 
\end{center}
\caption{
{\bf (a)} Survival probability $S_g(t|\x_0)$ and {\bf (b)} PDF
$H_g(t|\x_0)$ for two generations $g = 5$ (circles) and $g = 10$
(solid line) of the Koch snowflake with $\Theta =\pi/3$, $L = 2$, $D =
1$, $r_0 = 10^{-4}$ and $\theta_0 = \Theta/2$, starting near the angle
$\pi/3$ (Fig. \ref{fig:koch_g10}(c)), obtained by Monte Carlo
simulations with $N = 10^8$ particles.  Thin lines show the survival
probabilities $S_{\pi/3,\infty}(t|\x_0)$ and
$S_{2\pi/3,\infty}(t|\x_0)$ for the wedges of angles $\pi/3$ (dashed
line) and $2\pi/3$ (dash-dotted line) on panel {\bf (a)}, and the
associated PDFs $H_{\pi/3,\infty}(t|\x_0)$ and
$H_{2\pi/3,\infty}(t|\x_0)$ on panel {\bf (b)}.  }
\label{fig:St_simu}
\end{figure}

We first consider the starting point $\x_0$ close to a vertex of angle
$\pi/3$ on the boundary.  For convenience, we keep using local polar
coordinates, $\x_0 = (r_0,\theta_0)$, which are centered at the chosen
vertex of the boundary.  We set $r_0 = 10^{-4}$ and $\theta_0 =
\pi/6$, i.e., the starting point $\x_0$ lies on the ray in the middle
of the vertex, as illustrated on Fig. \ref{fig:koch_g10}(c).  The
survival probability $S_g(t|\x_0)$ for two generations $g = 5$ and $g
= 10$, as well as $S_{\pi/3,\infty}(t|\x_0)$ and
$S_{2\pi/3,\infty}(t|\x_0)$, are shown in Fig. \ref{fig:St_simu}(a).
For $g = 5$, the length $\ell_5 = L(1/3)^5 \approx 8.2\cdot 10^{-3}$
of the segments considerably exceeds the starting distance $h_0 = r_0
\sin(\pi/6) = 0.5\cdot 10^{-4}$ to the boundary.  As a consequence,
the survival probability $S_5(t|\x_0)$ is very close to that of the
wedge of angle $\pi/3$, as confirmed by the figure, with the
persistence exponent $\alpha = \pi/(2\Theta) = 3/2$.  In contrast, for
$g = 10$, one has $\ell_{10} = L(1/3)^{10} \approx 3.4\cdot 10^{-5}$,
which is much smaller than the initial distance.  As the inequalities
(\ref{eq:cond}) are now satisfied, one can expect the emergence of a
nontrivial intermediate regime.  Indeed, we observe deviations of the
survival probability $S_{10}(t|\x_0)$ from its lower bound
$S_{\pi/3,R}(t|\x_0) \approx S_{\pi/3,\infty}(t|\x_0)$.  Note also
that the upper bound given by the survival probability
$S_{2\pi/3,\infty}(t|\x_0)$ (with the persistence exponent
$\pi/(2\Theta') = 3/4$) is not tight.  Fitting $S_{10}(t|\x_0)$ by a
power law over an intermediate range of times gives the persistence
exponent close to $1.25$.  This value is smaller than the persistence
exponent $3/2$ of the wedge, but larger than the persistence exponent
$d_f/2 \approx 0.63$ predicted in \cite{Levitz06} (see further
discussions in Sec. \ref{sec:local}).  The associated PDFs are
illustrated on Fig. \ref{fig:St_simu}(b).  We emphasize on a very
broad range of relevant timescales, ranging from $t_{\rm mp}
\approx h_0^2/(6D) \approx 4\cdot 10^{-10}$ to $t_c
\approx 10^{-1}$, see Eq. (\ref{eq:t2}).  Note that the exponential
cut-off determined by $t_c$, is not shown here due to insufficient
accuracy of the empirical statistics for very large FPTs.

\begin{figure}
\begin{center}
\includegraphics[width=88mm]{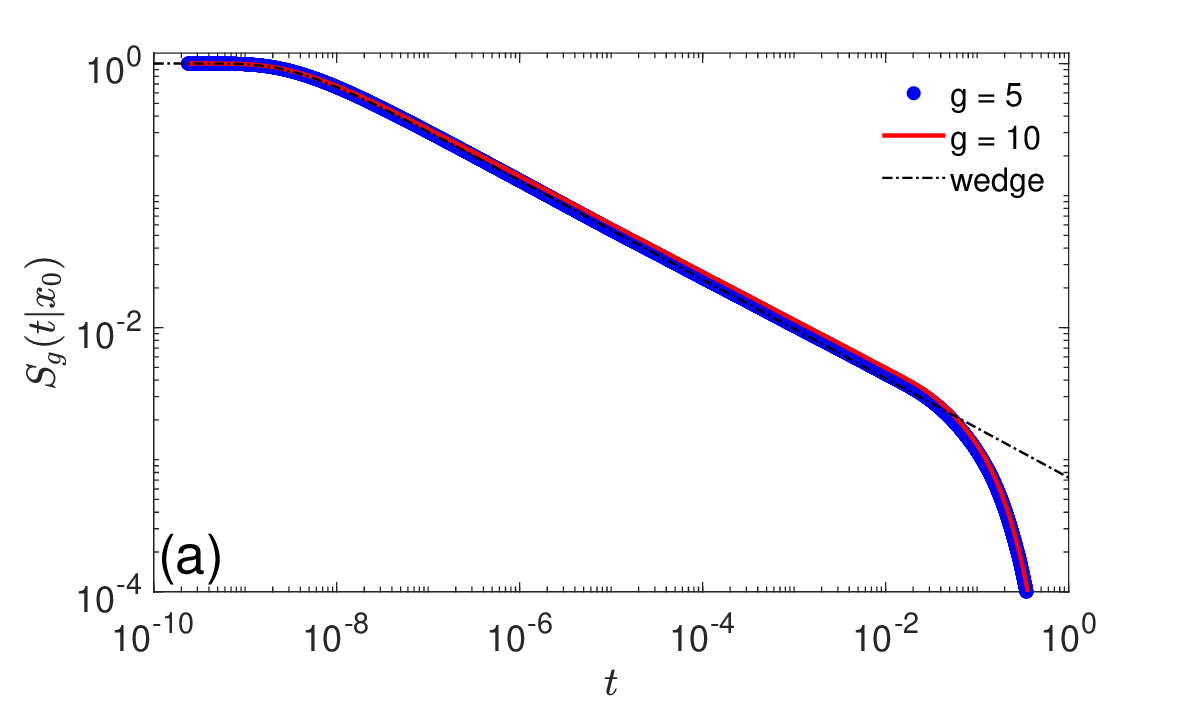} 
\includegraphics[width=88mm]{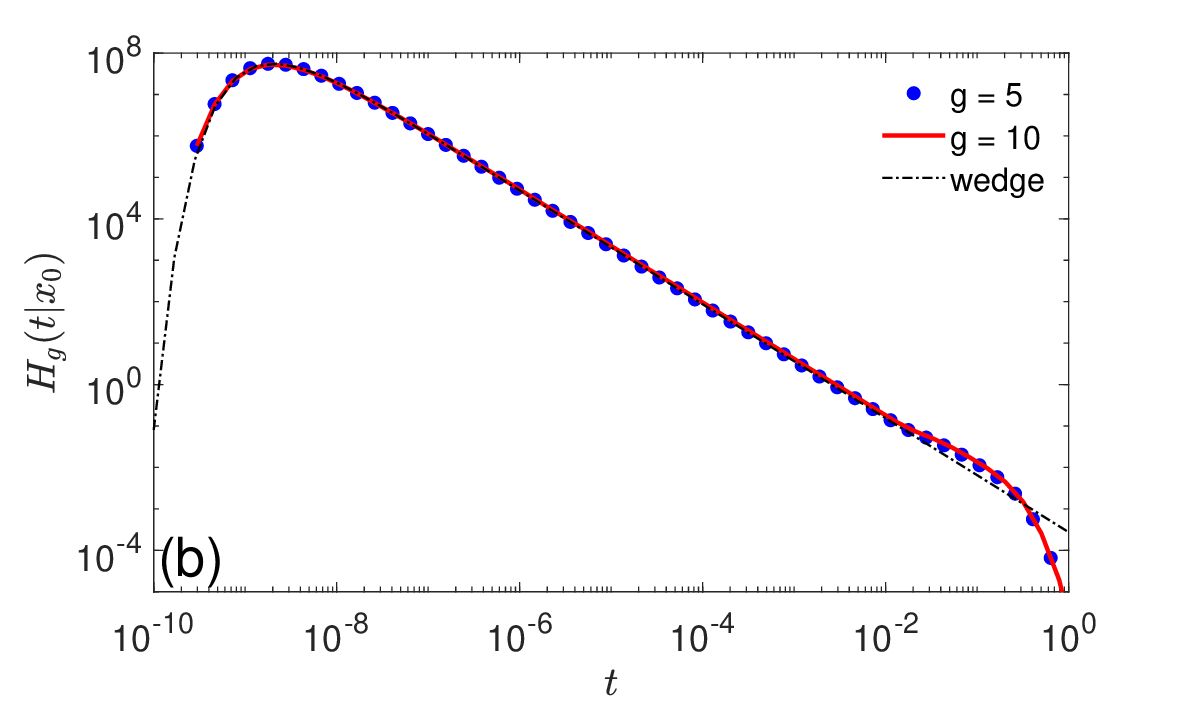} 
\end{center}
\caption{
{\bf (a)} Survival probability $S_g(t|\x_0)$ and {\bf (b)} PDF
$H_g(t|\x_0)$ for two generations $g = 5$ (circles) and $g = 10$
(solid line) of the Koch snowflake, with $\Theta =\pi/3$, $L = 2$, $D
= 1$, $r_0 = 10^{-4}$ and $\theta_0 = \Theta/2$, starting near the
angle $4\pi/3$ (Fig. \ref{fig:koch_g10}(b)), obtained by Monte Carlo
simulations with $N = 10^8$ particles.  Thin dash-dotted line shows
the survival probability $S_{4\pi/3,\infty}(t|\x_0)$ on panel {\bf
(a)} and the PDF $H_{4\pi/3,\infty}(t|\x_0)$ on panel {\bf (b)} for
the wedge of angle $4\pi/3$, which decay as $t^{-\alpha}$ and
$t^{-\alpha-1}$, respectively, with $\alpha = \pi/(2\cdot 4\pi/3) =
0.375$.  Note that the exponential decay starts at $t_c \approx 0.1$
given by Eq. (\ref{eq:t2}). }
\label{fig:St_simu2}
\end{figure}

The behavior is quite different for the starting point near the angle
$4\pi/3$, illustrated on Fig. \ref{fig:koch_g10}(b).  Figure
\ref{fig:St_simu2}(a) indicates that the survival probabilities
$S_5(t|\x_0)$ and $S_{10}(t|\x_0)$ are close to each other and to the
survival probability $S_{4\pi/3,\infty}(t|\x_0)$ for the wedge of
angle $4\pi/3$, which exhibits the power-law decay with the
persistence exponent $\alpha = \pi/(2\Theta) = 3/8$.  Quite
remarkably, $S_{4\pi/3,\infty}(t|\x_0)$ turns out to be an accurate
approximation for a broad range of times, up to $10^{-1}$.  At this
timescale, the particle starts to feel the confinement, as witnessed
by the expected exponential cut-off.  A similar behavior is found for
the PDF $H_g(t|\x_0)$ shown on Fig. \ref{fig:St_simu2}(b).

We conclude that the power-law decay of the survival probability near
the Koch snowflake strongly depends on the location of the starting
point; in particular, the persistence exponent can considerably vary,
e.g., by taking a value of $1.25$ in the vicinity of the angle $\pi/3$
and $0.38$ near the angle $4\pi/3$.  How is it possible that the
long-time behavior of the survival probability is so sensitive to the
starting point?  Indeed, large FPTs correspond to the random
trajectories that moved away from the boundary (to avoid absorption at
short times) and are thus supposed to ``forget'' about their starting
point.
This behavior may sound even more puzzling after recalling the former
study by Levitz {\it et al.}, in which the starting point was
uniformly distributed near the absorbing boundary \cite{Levitz06}.  In
this work, the persistence exponent $\alpha$ was related via
Eq. (\ref{eq:S_fractal}) to the fractal dimension $d_f$ of the
boundary.  Can one claim that the power-law decay $t^{-d_f/2}$ is an
average of different power laws with $\x_0$-dependent persistence
exponents over the starting point $\x_0$?  In the next subsection, we
refine the analysis of the power-law decay in order to clarify this
puzzling behavior.

\subsection{Local persistence exponent}
\label{sec:local}

The persistence exponent $\alpha$ is usually estimated by fitting the
survival probability to a power law.  However, as the power-law
behavior presents an intermediate regime in our setting, one needs to
choose properly the range of timescales for fitting.  As a
consequence, the fitting procedure may be inaccurate and biased.  To
avoid these issues and to get a finer insight onto the asymptotic
behavior of the survival probability, we introduce {\it the local
persistence exponent} as the (negative) logarithmic derivative of
$S_g(t|\x_0)$:
\begin{equation}
\alpha(t) = - \frac{d\ln S_g(t|\x_0)}{d\ln t} \,.
\end{equation}
If the survival probability decayed as a power law with an exponent
$\alpha$ at all times, one would simply get $\alpha(t) = \alpha$.
However, the actual behavior of the survival probability is more
sophisticated.  At very short times, one has $1 - S_g(t|\x_0) \propto
e^{-h_0^2/(4Dt)}$, where $h_0$ is the distance to the boundary, so
that $\alpha(t) \to 0$ as $t\to 0$.  At very long times, one has
$S_g(t|\x_0) \propto e^{-Dt\lambda_{\rm min}}$ so that $\alpha(t)
\approx D\lambda_{\rm min} t \to +\infty$ because the exponential
decay is faster than any power law.  But if there was an intermediate
power-law regime in between, one would get a plateau of $\alpha(t)$
with a constant value that can associated with the persistence
exponent.

To evaluate the local persistence exponent numerically from a given
empirical CDF, we define a range of $K+1$ times $t_k = t_0
(t_K/t_0)^{k/K}$ (with $k = 0,1,\ldots,K$), which are equally spaced
on the logarithmic scale between prescribed bounds $t_0$ and $t_K$.
By interpolating $S_g(t_k|\x_0)$ from the empirical CDF, we set
\begin{equation}
\alpha(t'_k) = - \frac{\ln(S_g(t_{k}|\x_0)/S_g(t_{k-1}|\x_0))}{\ln (t_{k}/t_{k-1})}  \quad (k = 1,\ldots ,K),
\end{equation}
estimated at the intermediate time $t'_k = \sqrt{t_k t_{k-1}}$ between
$t_{k-1}$ and $t_k$ (the geometric mean is used to respect the
logarithmic spacing of points).  For $K$ large enough, these values
provide an accurate approximation for the logarithmic derivative.  For
the considered examples, we choose $K = 100$.  We checked that the
obtained results did not almost change when $K$ was doubled.  We also
note that additional simulations with a minimal threshold $T_{\rm
min}$ were needed to get more accurate estimates of $\alpha(t)$ at
long times (see Sec. \ref{sec:numerics}).

\begin{figure}
\begin{center}
\includegraphics[width=88mm]{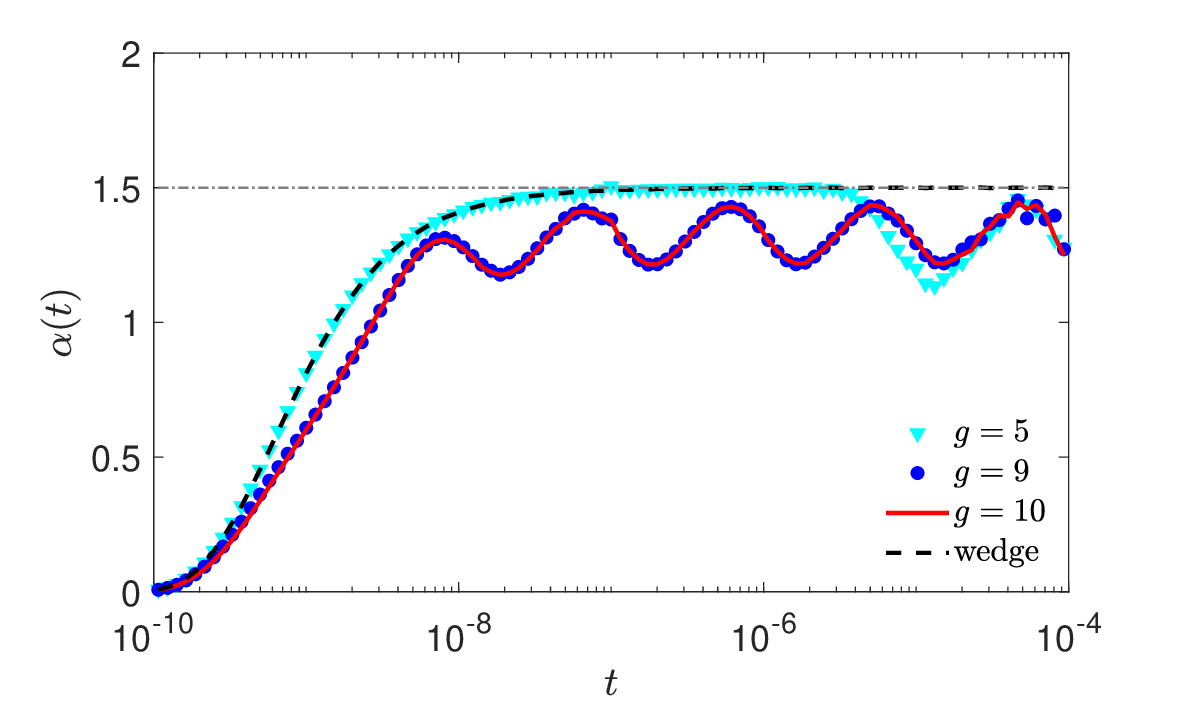} 
\end{center}
\caption{
Local persistence exponent $\alpha(t)$ for three generations $g = 5$
(triangles), $g = 9$ (circles) and $g = 10$ (solid line) of the Koch
snowflake, with $\Theta =\pi/3$, $L = 2$, $D = 1$, $r_0 = 10^{-4}$ and
$\theta_0 = \Theta/2$, starting near the angle $\pi/3$
(Fig. \ref{fig:koch_g10}(c)), obtained by Monte Carlo simulations with
$N = 10^8$ particles (its behavior for $t > T_{\rm min} = 10^{-7}$ was
estimated from additional simulations, see Sec. \ref{sec:numerics}).
Dashed line presents the local persistence exponent for the wedge of
angle $\pi/3$, whereas dash-dotted horizontal line indicates its
persistence exponent $3/2$ (in the limit $t\to \infty$).}
\label{fig:alpha_pi3}
%
\end{figure}

Figure \ref{fig:alpha_pi3} illustrates the local persistence
exponents, which were obtained from the survival probabilities shown
on Fig. \ref{fig:St_simu}(a), corresponding to the starting point near
the vertex of angle $\pi/3$ (Fig. \ref{fig:koch_g10}(c)).  Due to
rarity of too small and too large FPTs, we could not access
$\alpha(t)$ accurately at very short and very long times so that we
focused on the timespan between $t_0 = 10^{-10}$ and $t_K = 10^{-4}$.
As this figure conveys one of our main results, we discuss it in
detail.

Let us first examine the local persistence exponent for the wedge of
angle $\Theta = \pi/3$ (dashed line).  To get this curve, we first
computed the survival probability $S_{\Theta,\infty}(t|\x_0)$ from
Eq. (\ref{eq:St_wedge}) at numerous time instances, recasted it as an
empirical CDF, and then applied the above numerical procedure.
Expectedly, $\alpha(t)$ starts from $0$ at $t\to 0$ and then grows
monotonically to reach a plateau for $t \gtrsim 10^{-7}$.  This
plateau corresponds to the persistence exponent $\alpha =
\pi/(2\Theta) = 3/2$ of the wedge.
Next, we look at $\alpha(t)$ for the generation $g=5$ (triangles).  As
the survival probability $S_5(t|\x_0)$ was very close to
$S_{\Theta,\infty}(t|\x_0)$ up to $t \simeq 4\cdot 10^{-6}$
(Fig. \ref{fig:St_simu}(a)), their local persistence exponents are
almost identical over this time range.  However, at longer times,
$\alpha(t)$ starts to deviate from the plateau.  This timescale
determines a typical diffusion length $\sqrt{4Dt} \approx 4\cdot
10^{-3}$, which is comparable to the length $\ell_5 \approx 8.2\cdot
10^{-3}$ of the segments.  In other words, at times $t \gtrsim 4\cdot
10^{-6}$, the particle starts to ``feel'' that the absorbing boundary
is not a wedge but the 5-th generation of the Koch snowflake, and thus
the survival probability $S_g(t|\x_0)$ deviates from
$S_{\Theta,\infty}(t|\x_0)$.

The most striking behavior is observed for generations $g=9$ and
$g=10$.  After a short transient regime of growing $\alpha(t)$, the
local persistence exponent starts to {\it oscillate} (on the
logarithmic scale), still remaining below $3/2$.  Since the curves for
generations $g=9$ and $g=10$ are almost identical, we can claim that
this behavior is independent of the generation (if $g$ is not small)
and thus representative of the behavior in the limit $g\to\infty$,
i.e., to the fractal boundary.  The earlier reported value of the
persistence exponent from fitting a power law, $1.25$, is the average
level of oscillations.  Note that the local persistence exponent for
the generation $g = 5$ also approaches these log-periodic oscillations
at times $t \gtrsim 4\cdot 10^{-6}$.  This numerical observation
conveys that the survival probability in the Koch snowflake does not
exhibit a power-law decay and thus cannot be characterized by a
persistence exponent.  In turn, the local persistence exponent reveals
much finer details of the asymptotic behavior and its relation to the
geometric structure of the fractal boundary.

How many log-periodic oscillations of $\alpha(t)$ can exist?  Figure
\ref{fig:alpha_pi3} revealed five log-periodic oscillations on the
considered timespan from $10^{-10}$ to $10^{-4}$.  Extending this
pattern hypothetically to longer times, several additional
oscillations can be expected, up to the exponential cut-off $t_c
\approx 10^{-1}$, above which $\alpha(t)$ should exhibit a linear
growth with $t$.  In other words, the cut-off time $t_c$ sets the
upper bound on the time range of log-periodic oscillations.  In turn,
the lower bound is determined by the transient growth of $\alpha(t)$
at short times, which is controlled by the initial distance to the
boundary.  If one could generate a sufficiently accurate statistics of
the FPT at arbitrarily small distance $h_0$ from the boundary $\pa_g$
of arbitrary high generation $g$, the log-periodic pattern of
$\alpha(t)$ could be extended to the left to arbitrarily short times.

The observed log-periodic oscillations reflect self-similarity of the
Koch boundary.  Indeed, if $T_1, T_2, \ldots$ denote the time
instances of the successive maxima of $\alpha(t)$, we observe
$T_{k+1}/T_k \approx 9$, whereas the ratio of the associated length
scales, $\sqrt{DT_{k+1}}/\sqrt{DT_k}$, is around $3$.  This is
precisely the scaling factor between two successive generations of the
Koch boundary with $\Theta = \pi/3$.  This simple relation suggests
the following interpretation.

\begin{figure}
\begin{center}
\includegraphics[width=42mm]{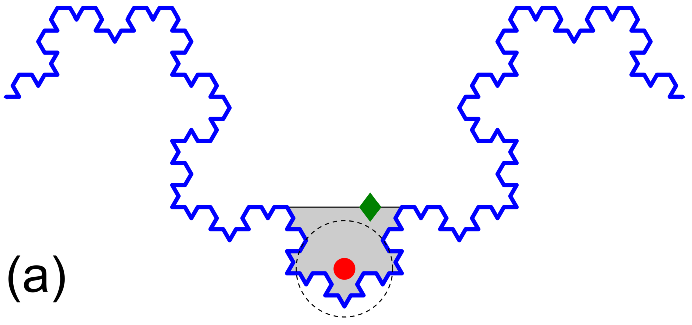} 
\includegraphics[width=42mm]{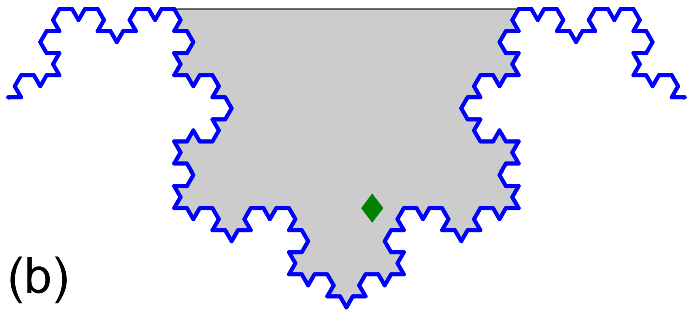} 
\end{center}
\caption{
{\bf (a)} Local environment around the starting point (red circle)
located near the boundary of the Koch snowflake with $\Theta =\pi/3$
(a small fraction of the boundary is shown).  Dashed circle of radius
$h$ indicates a close vicinity of the boundary, on which most
particles are rapidly absorbed.}  To avoid rapid absorption, the
particle needs to leave the gray zone, e.g., through a randomly chosen
point indicated by green diamond.  {\bf (b)} To avoid absorption at a
longer time, the particle at green diamond needs to leave a new gray
zone, which is three times bigger than the previous one.
\label{fig:koch_cell}
\end{figure}

Let us look at the local geometric environment near the starting point
indicated by a gray zone in Fig. \ref{fig:koch_cell}(a).  Most
particles are absorbed by the boundary in a close vicinity of the
starting point, within a distance $h$, which is comparable to the
distance $h_0$ to the boundary, yielding the most probable FPT of the
order of $h^2/(6D)$.  As a consequence, the short-time behavior of the
survival probability is mostly determined by this local environment.
Moreover, since the local persistence exponents for generations $g =
9$ and $g = 10$ on Fig. \ref{fig:alpha_pi3} were almost identical, the
inclusion of the smaller geometric details at the iteration from $g =
9$ to $g = 10$ does not seem to affect the first-passage times.  This
is consistent with the fact that most trajectories are stopped at the
most accessible parts of the boundary (see
\cite{Grebenkov05a,Grebenkov05b} for the related discussion on the
harmonic measure).
In turn, the survival probability at longer times describes the
trajectories that managed to avoid hitting this local environment and
thus reached a larger gray zone shown in Fig. \ref{fig:koch_cell}(b).
Qualitatively, such trajectories can be treated as if they just
started from a random point inside the larger gray zone.  The
self-similar structure of the boundary allows one to repeat this
argument again and again so that the survival of the particle at
longer and longer times can be understood as a sequence of successful
``escapes'' from one zone to the larger one.  This sequence is
repeated until the particle arrives into central part of the Koch
snowflake where it starts to ``feel'' the confinement and thus
switches to the exponential decay.  As a consequence, the log-periodic
pattern in the local persistence exponent $\alpha(t)$ reflects the
self-similarity of the absorbing boundary.  Qualitatively, the number
of oscillations should thus be equal to the number of ``escapes'',
which is a purely geometric characteristic that depends on the
starting point.

\begin{figure}
\begin{center}
\includegraphics[width=88mm]{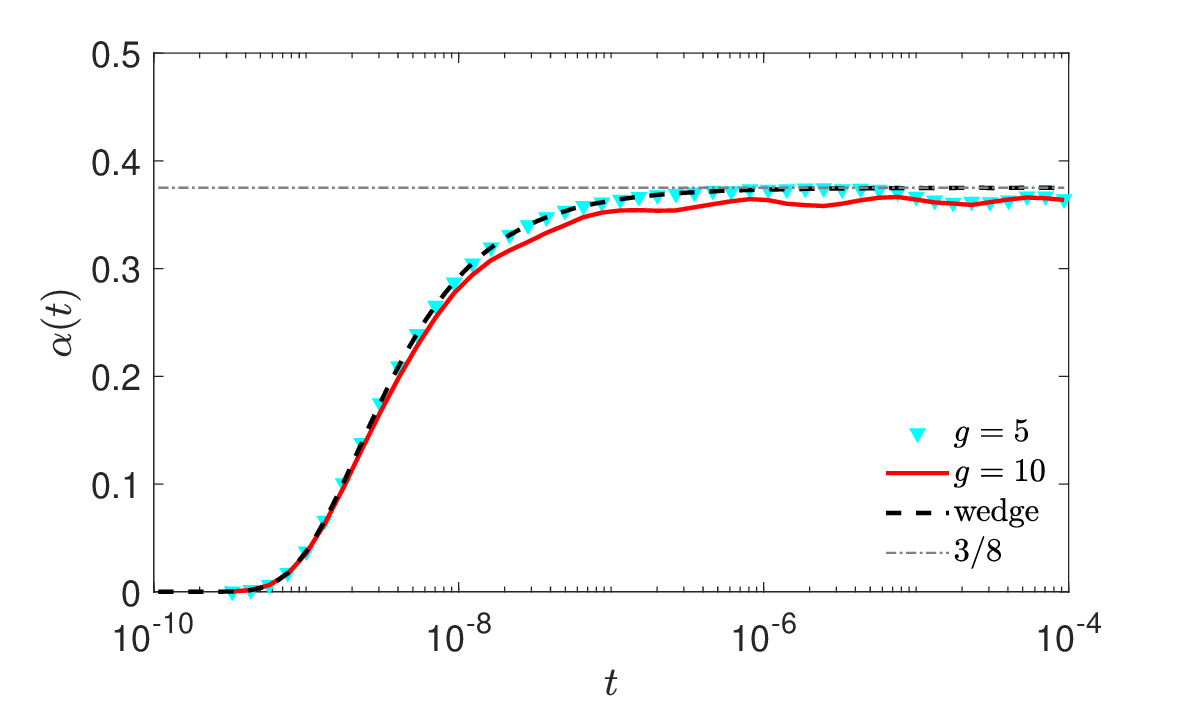} 
\end{center}
\caption{
Local persistence exponent $\alpha(t)$ for two generations $g = 5$
(triangles) and $g = 10$ (solid line) of the Koch snowflake, with
$\Theta =\pi/3$, $L = 2$, $D = 1$, $r_0 = 10^{-4}$ and $\theta_0 =
2\pi/3$, starting near the angle $4\pi/3$
(Fig. \ref{fig:koch_g10}(b)), obtained by Monte Carlo simulations with
$N = 10^8$ particles.  Dashed line shows the local persistence
exponent of the survival probability $S_{4\pi/3,\infty}(t|\x_0)$ of
the wedge of angle $4\pi/3$, while dash-dotted horizontal line
indicates its long-time asymptotic value $3/8$.}
\label{fig:alpha_pi3_4pi3}
\end{figure}

After this qualitative argument, it is instructive to revisit the
survival probability shown on Fig. \ref{fig:St_simu2}(a) for another
starting point near the angle $4\pi/3$.  Figure
\ref{fig:alpha_pi3_4pi3} shows that the local persistence exponents
for generations $g = 5$ and $g = 10$ are almost identical and are
actually close to that of the wedge of angle $4\pi/3$.  At long times,
the latter reaches the limit $3/8$, i.e., the persistence exponent for
the wedge.  In this setting, the diffusing particle is either rapidly
absorbed in the local environment of its starting point (which is
close to that of a wedge), or easily reaches the central part of the
Koch snowflake, with almost no effect of the self-similar boundary.
In other words, there is no a sequence of ``escapes'' from smaller to
larger zones, and the local persistence exponent $\alpha(t)$ remains
almost constant (and close to the value $3/8$ for the wedge).

\begin{figure}
\begin{center}
\includegraphics[width=88mm]{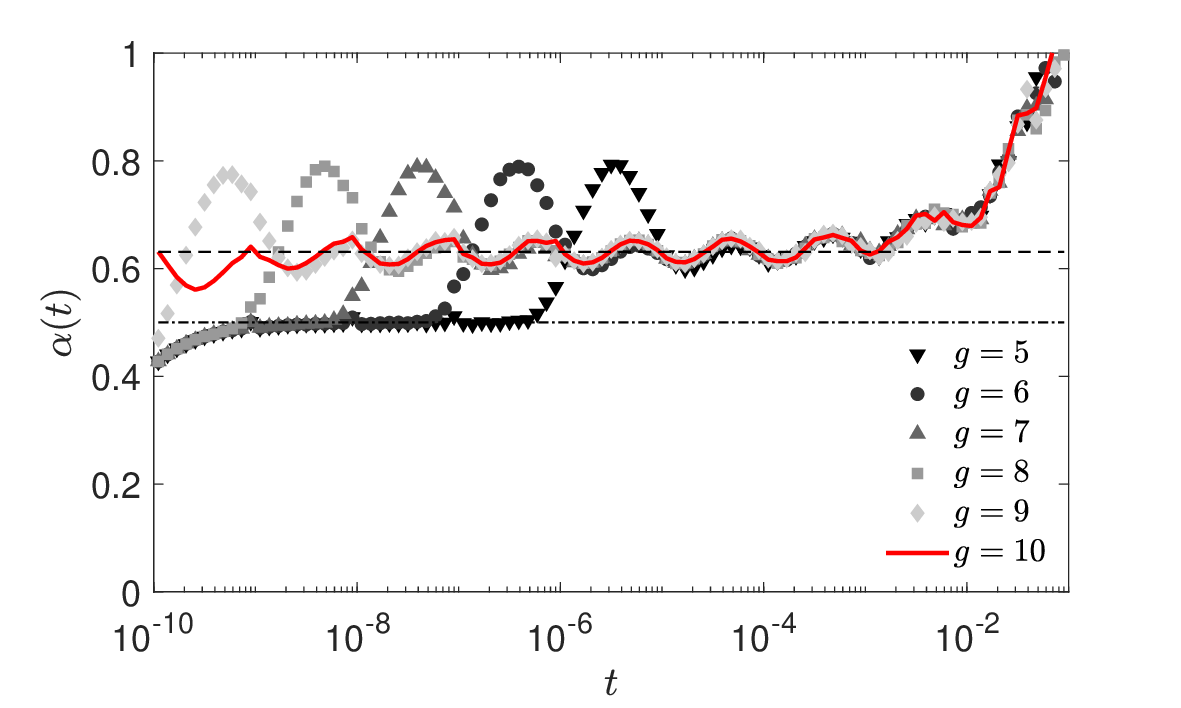} 
\end{center}
\caption{
Local persistence exponent $\alpha(t)$ for six generations from $g =
5$ (light diamonds) to $g = 10$ (solid line) of the Koch snowflake,
with $\Theta =\pi/3$, $L = 2$, $D = 1$, and the uniformly distributed
starting point at distance $h_0 = 10^{-5}$ from the boundary, obtained
by Monte Carlo simulations with $N = 10^8$ particles (an extended
simulation for $t > T_{\rm min} = 10^{-6}$ with $N_0 = 10^7$ was also
included, see Sec. \ref{sec:numerics}).  Dashed horizontal line
indicates the persistence exponent $d_f/2 \approx 0.63$, while
dash-dotted line presents the persistence exponent $1/2$ of a flat
boundary.}
\label{fig:alpha_pi3_uniform}
\end{figure}

Finally, we revisit the asymptotic behavior of the survival
probability $S_g(t)$ in the case when the starting point is uniformly
distributed near the boundary.  As discussed earlier, this survival
probability exhibits a power-law decay (\ref{eq:S_fractal}) with the
persistence exponent $\alpha = d_f/2$ \cite{Levitz06}.  Figure
\ref{fig:alpha_pi3_uniform} shows the local persistence exponent in
this case for generations from $g = 5$ to $g = 10$.  For $g = 10$, we
get small oscillations around $\alpha = d_f/2 \approx 0.63$.  Since
the chosen distance $h_0 = 10^{-5}$ to the boundary is smaller than
the segment length $\ell_{10} \approx 3.39\cdot 10^{-5}$, we do not
see a transient regime, which should be present at much smaller times.
For $g = 5$, one has $h_0 \ll \ell_5 \approx 8.2\cdot 10^{-3}$ so that
the particle first explores a flat local environment, and the local
persistence exponent is close to $1/2$ at short times.  In turn, one
observes a transition to $\alpha \approx 0.63$ at longer times.  As
the generation $g$ increases, $\ell_g$ gets smaller and smaller, so
that the transition to $\alpha \approx 0.63$ occurs at shorter and
shorter times.  For all considered generations, we also observe a
growth of $\alpha(t)$ at times $t$ larger than $10^{-2}$, as expected
at very long times.  This observation was possible because the
survival probability $S_g(t)$ decays slower than $S_g(t|\x_0)$
considered earlier, so that our Monte Carlo simulations were accurate
enough to estimate $S_g(t)$ over a broader timespan.

\section{Discussion and conclusion}
\label{sec:conclusion}

In this paper, we studied the distribution of the FPT to the absorbing
self-similar boundary of the Koch snowflake.  An accurate examination
of this problem in a circular sector and a wedge allowed us to outline
three distinct regimes at short, intermediate, and long times.  The
first and the third regimes characterize respectively too short and
too long trajectories and exhibit universal, well-known features.  In
turn, the intermediate regime is geometry-dependent and less
understood in general.  When the starting point is fixed and located
near a smooth absorbing boundary, the intermediate regime exhibits a
power-law decay with the persistence exponent $\alpha = 1/2$,
inherited from the L\'evy-Smirnov distribution on the half-line.  For
irregular boundaries, however, the persistence exponent is controlled
by singularities (e.g., $\alpha = \pi/(2\Theta)$ for a wedge of angle
$\Theta$).  If there are many different angles, their effect onto the
distribution of the FPT can be more sophisticated.  One can expect
that the angle that is the nearest to the starting point would affect
the survival probability first, at time scales of the order $r_0^2/D$,
where $r_0$ is the distance to that angle.  However, at longer times,
the survived random trajectories can explore more distant regions and
thus be affected by farther angles.  This intuitive picture gives an
idea that the decay of the survival probability is not necessarily
controlled by a single persistence exponent but may exhibit distinct
decay laws at different time scales.  In other words, a single power
law may be insufficient to fully characterize the survival of a
particle in complex domains.

Our extensive Monte Carlo simulations confirmed this intuitive
picture.  For this purpose, we introduced the local persistence
exponent $\alpha(t)$ as the logarithmic derivative of the survival
probability, i.e., $S(t) \propto t^{-\alpha(t)}$.  As this relation
can formally be applied to any positive function $S(t)$, the local
persistence exponent $\alpha(t)$ is just a convenient representation
of the survival probability.  In particular, if $\alpha(t)$ varies
with time, the survival probability does not obey a power law.  Our
numerical results revealed four possible regimes for $\alpha(t)$:
(i) $\alpha(t) \to 0$ as $t\to 0$ in agreement with the L\'evy-Smirnov
short-time behavior of the survival probability; in particular, there
is a transient growth of $\alpha(t)$ at short times;
(ii) $\alpha(t) \approx 1/2$ or $\alpha(t) \approx \pi/(2\Theta)$ at
early intermediate times when $h_0 \ll \ell_g$, i.e., if the particle
started close to a linear segment of length $\ell_g$ of the polygonal
boundary $\pa_g$ (the $g$-th generation of the Koch snowflake);
(iii) $\alpha(t)$ exhibits log-periodic oscillations at late
intermediate times; and
(iv) $\alpha(t) \propto t$ at long times due to the exponential decay
of the survival probability in a bounded domain.  
While the first and the last regimes are universal and always present
for any bounded domain, the two intermediate regimes strongly depend
on the considered domain and the starting point.  For instance, in the
case of finite generations of the Koch snowflake, we gave examples
when both regimes are present (curve for $g = 5$ on
Fig. \ref{fig:alpha_pi3}), only the second regime is present
(Fig. \ref{fig:alpha_pi3_4pi3}), only the third regime is present
(curve for $g = 10$ on Fig. \ref{fig:alpha_pi3}).

The log-periodic oscillations of the local persistence exponent of the
survival probability present the main finding of the paper.  The
self-similarity of the absorbing boundary was identified as their
origin.  Indeed, the survival at longer and longer times requires
moving away from the absorbing boundary, as far as possible, into the
bulk.  The repeated pattern of $\alpha(t)$ results from a sequence of
successive ``escapes'' into larger and larger regions, which are
rescaled copies of each other.  The number of log-periodic
oscillations reflects how many such successively enlarged regions are
present on the path from the starting point to the central part of the
Koch snowflake.  In particular, in the case of the infinite generation
of the fractal boundary, one should be able to get any number of
log-periodic oscillations by locating the starting point arbitrarily
close to the boundary.  In other words, the oscillatory behavior of
$\alpha(t)$ can cover an arbitrarily broad timespan.  In this regime,
the survival probability does not exhibit a power-law decay and thus
cannot be characterized by a single persistence exponent.  This
important feature of the diffusive dynamics seems to be overseen in
the past, possibly because most earlier works were focused either on a
uniform starting point, or on the MFPT.  Fixing the starting point
near the absorbing boundary allowed us to uncover new and richer
aspects of this phenomenon.

This numerical discovery raises many fundamental questions.  While
log-periodicity of $\alpha(t)$ can be explained by self-similarity,
the shape of its motive remains unclear.  As the diffusing particle
aims to escape from the local polygonal environment formed by a
mixture of different angles (here, $\pi/3$ and $4\pi/3$), one can
speculate that the shape of one period in $\alpha(t)$ results from
averaging between distinct power laws over a limited timespan.  It
would be interesting to relate the shape of the periodic pattern to
the geometric structure of the local environment.  Moreover, one can
also determine how the average $\alpha(t)$ and the amplitude of its
oscillations depend on the geometry (e.g., on the angle $\Theta$ of
the generator of the Koch snowflake or, more generally, on the fractal
dimension).  Another question concerns the analysis of more
sophisticated fractal boundaries.  In fact, even though fractal
boundaries often emerge in natural applications
\cite{Mandelbrot,Sapoval,Gouyet,benAvraham}, their self-similarity is
generally not perfect, and one often deals with heterogeneous
structures, multifractals, or random fractals.  For instance, one can
still consider the iterative construction of the Koch snowflake but
choose randomly the orientation of the generator.  In this way, one
would produce a random fractal with the same fractal dimension $d_f$
given by Eq. (\ref{eq:df}).  The earlier numerical work by Filoche and
Sapoval showed that the steady-state diffusive fluxes towards random
and deterministic fractals with the same $d_f$ are essentially
identical, i.e., randomness is not so relevant \cite{Filoche00}.  What
would be the impact of randomness or heterogeneity onto the asymptotic
behavior of the survival probability?  Another interesting perspective
consists in extending the above analysis to exterior problems (i.e.,
diffusion outside a target with a fractal boundary) and to higher
dimensions.

Finally, the survival probability in any bounded domain admits a
spectral decomposition over the eigenvalues $\lambda_n$ and
eigenfunctions $u_n(\x)$ of the Dirichlet Laplace operator in that
domain:
\begin{equation}
S(t|\x_0) = \sum\limits_{n=1}^\infty e^{-Dt\lambda_n} u_n(\x_0) \int\limits_\Omega d\x \, u_n(\x). 
\end{equation}
The log-periodic pattern of the local persistence exponent of
$S(t|\x_0)$ should thus be related to the geometrical structure of
these eigenmodes, which often exhibit self-similar features
\cite{Sapoval91,Lapidus95,Even99,Daudert07,Banjai07,Grebenkov13}.
A similar relation between the log-periodic behavior of the heat
kernel and the spectrum of the Laplace operator was investigated for
{\it fractal domains} (see \cite{Akkermans09} and references therein).
In particular, the MFPT in such domains was shown to exhibit spatial
log-periodic oscillations \cite{Akkermans12}.  More generally,
log-periodic oscillations were observed in various physical systems
exhibiting discrete scale invariance, with examples ranging from
critical phenomena and chaotic dynamics to turbulence and financial
crashes \cite{Sornette98,Cagnetta15,Luck24}.
As explained in Sec. \ref{sec:intro}, our setting of ordinary
diffusion in a Euclidean domain with a {\it fractal boundary} is
drastically different and, perhaps, more challenging for mathematical
analysis.  For instance, the fractal boundary is known to affect only
the subleading term of the heat trace and the Weyl's law
\cite{Brossard86,Lapidus91,Lapidus96} (see also
\cite{Zhang10} for random walks on Vicsek fractals), i.e., it requires
finer mathematical tools.  In particular, we saw the strong dependence
of the asymptotic behavior on the starting point.  Further
clarifications of the fundamental relation between log-periodic
oscillations of the survival probability and the Laplacian spectrum
presents an important perspective of this study.

\begin{acknowledgments}
The authors thank Prof. M. Dolgushev for reading the manuscript and
fruitful discussions.  D.S.G. acknowledges the Simons Foundation for
supporting his sabbatical sojourn in 2024 at the CRM (University of
Montr\'eal, Canada), as well as the Alexander von Humboldt Foundation
for support within a Bessel Prize award.
\end{acknowledgments}


\appendix
\section{Computation of the second moment}
\label{sec:T2}

Using the spectral representation (\ref{eq:St_sector}), one can
evaluate the second moment of the FPT in the sector:
\begin{widetext}
\begin{align*}
T^{(2)}_{\Theta,R} & = \int\limits_0^\infty dt\, t^2 \, H_{\Theta,R}(t|\x_0) 
= 2\int\limits_0^\infty dt\, t\, S_{\Theta,R}(t|\x_0) \\
& = \frac{8}{\Theta} \sum\limits_{n=1}^\infty \sum\limits_{k=1}^\infty \frac{c_{nk}^2}{D^2\lambda_{nk}^2} \sin(\nu_n \theta_0) 
J_{\nu_n}(\alpha_{nk} r_0/R) \frac{1-(-1)^n}{\nu_n} R^2 \int\limits_0^1 dx \, x \, J_{\nu_n}(\alpha_{nk} x) \\
& = \frac{8R^4}{\Theta D^2} \sum\limits_{n=1}^\infty \frac{1-(-1)^n}{\nu_n} \sin(\nu_n \theta_0) 
\int\limits_0^1 dx \, x \, \sum\limits_{k=1}^\infty 
\frac{J_{\nu_n}(\alpha_{nk} r_0/R)   J_{\nu_n}(\alpha_{nk} x)}{\alpha_{nk}^4 J_{\nu_n+1}^2(\alpha_{nk})} \,.
\end{align*}
\end{widetext}
In order to compute the last sum, we use the summation identity (see
(D10) from Table 3 in \cite{Grebenkov21}):
\begin{align} \nonumber
& \sum\limits_{k=1}^\infty \frac{2J_{\nu_n}(\alpha_{nk} x_0) J_{\nu_n}(\alpha_{nk} x)}{(z^2-\alpha_{nk}^2) J_{\nu_n+1}^2(\alpha_{nk})} \\
& = \frac{\pi}{2} \biggl(Y_{\nu_n}(zx_0) - J_{\nu_n}(zx_0) \frac{Y_{\nu_n}(z)}{J_{\nu_n}(z)}\biggr) J_{\nu_n}(zx) ,
\end{align}
which is valid for $0 \leq x \leq x_0 \leq 1$ (note that the factor
$\pi/2$ was missing in Eq. (D10)).  Evaluating the derivative with
respect to $z$, dividing by $z$ and taking the limit $z\to 0$, we get
for $0 \leq x \leq x_0 \leq 1$
\begin{align} \nonumber
& \sum\limits_{k=1}^\infty \frac{J_{\nu_n}(\alpha_{nk} x_0) J_{\nu_n}(\alpha_{nk} x)}{\alpha_{nk}^4 J_{\nu_n+1}^2(\alpha_{nk})}
= \frac{x^{\nu_n}}{16 \nu_n(\nu_n+1) x_0^{\nu_n}} \\
& \times \biggl(x_0^2 \bigl(x_0^{2\nu_n} - (1+a_{\nu_n}) x_0^{2\nu_n-2} + a_{\nu_n}\bigr) - x^2 \bigl(1 - x_0^{2\nu_n}\bigr)\biggr) ,
\end{align}  
where $a_\nu = (\nu+1)/(\nu-1)$.  Substituting this expression into
the above formula for the second moment, we get after simplifications
\begin{equation}
T^{(2)}_{\Theta,R} = \frac{R^4}{\Theta D^2} \sum\limits_{n=1}^\infty \frac{1-(-1)^n}{\nu_n(\nu_n+1)} \sin(\nu_n \theta_0) f_{\nu_n}(r_0/R),
\end{equation}
where
\begin{align} \nonumber
f_\nu(x_0) & = \frac{4(\nu+1)}{(\nu^2-4)(\nu^2-16)} x_0^4  \\   \label{eq:fnu}
&  -\frac{\nu+6}{(\nu+2)(\nu^2-16)} x_0^\nu + \frac{1}{\nu^2-4} x_0^{\nu+2} \,.
\end{align}   
Note that the sum involving the first term in Eq. (\ref{eq:fnu}) can
be evaluated exactly that yields
\begin{align}  \nonumber
T^{(2)}_{\Theta,R} & = T^{(2)}_{\Theta,\infty} + 
\frac{R^4}{\Theta D^2} \sum\limits_{n=1}^\infty \frac{1-(-1)^n}{\nu_n(\nu_n+1)} \sin(\nu_n \theta_0) \\  \label{eq:T2_sector}
& \times \biggl(  -\frac{(\nu_n+6) (r_0/R)^{\nu_n}}{(\nu_n+2)(\nu_n^2-16)}  + \frac{(r_0/R)^{\nu_n+2}}{\nu_n^2-4}   \biggr),
\end{align}
where
\begin{equation}  \label{eq:T2_Rinf}
T^{(2)}_{\Theta,\infty} = \frac{r_0^4}{96 D^2} \biggl(\frac{\cos(4\theta_0-2\Theta)}{\cos(2\Theta)} 
- 4 \frac{\cos(2\theta_0-\Theta)}{\cos(\Theta)} + 3 \biggr).
\end{equation}  
Combining this result with Eq. (\ref{eq:MFPT_sector}) for the first
moment, we get the standard deviation of the FPT,
\begin{equation}
\delta T_{\Theta,R} = \sqrt{T^{(2)}_{\Theta,R} - (T_{\Theta,R})^2} .
\end{equation}

In order to determine the leading order in the asymptotic behavior of
the second moment as $r_0\to 0$, we distinguish two cases.  When
$\Theta > \pi/4$, one has $\nu_1 = \pi/\Theta < 4$, so that the
leading-order contribution comes from the term $n = 1$ of the series
in Eq. (\ref{eq:T2_sector}) that scales as 
\begin{equation}
\frac{D^2}{R^4} T^{(2)}_{\Theta,R} \approx \frac{2(\nu_1+6)\sin(\nu_1\theta_0)}{\pi (\nu_1+1)(\nu_1+2)(16-\nu_1^2)} (r_0/R)^{\pi/\Theta} .
\end{equation}
In turn, if $\Theta < \pi/4$, the leading contribution is given by
$T^{(2)}_{\Theta,\infty}$ from Eq. (\ref{eq:T2_Rinf}), which scales as
$r_0^4$.  Combining these relations with Eqs. (\ref{eq:TTheta_asympt})
for the first moment, we deduce the asymptotic behavior of the
standard deviation:
\begin{widetext}
\begin{equation}  \label{eq:std_asympt}
\frac{D}{R^2} \delta T_{\Theta,R} \approx \begin{cases}  
\displaystyle \sqrt{\frac{2(\nu_1+6)\sin(\nu_1\theta_0)}{\pi (\nu_1+1)(\nu_1+2)(16-\nu_1^2)}} (r_0/R)^{\pi/(2\Theta)} 
 \hskip 53mm (\Theta > \pi/4), \cr
\displaystyle \frac{1}{\sqrt{96}}\biggl(\frac{\cos(4\theta_0-2\Theta)}{\cos(2\Theta)} 
- 6\frac{\cos^2(2\theta_0-\Theta)}{\cos^2(\Theta)} + 8 \frac{\cos(2\theta_0-\Theta)}{\cos(\Theta)} - 3 \biggr)^{1/2} (r_0/R)^2 
\qquad   (\Theta < \pi/4).   \end{cases}
\end{equation}
\end{widetext}
One can thus distinguish three regimes:
\begin{equation}  \label{eq:coeff_variation}
\frac{\delta T_{\Theta,R}}{T_{\Theta,R}} \propto \begin{cases}  1  \hskip 24.5mm (0< \Theta < \pi/4), \cr
(r_0/R)^{\pi/(2\Theta) - 2} \quad (\pi/4 < \Theta < \pi/2), \cr
(r_0/R)^{-\pi/(2\Theta)}  \hskip 5mm (\pi/2 < \Theta). \end{cases}
\end{equation}
In particular, one sees that the relative standard deviation diverges
as $r_0\to 0$ for $\Theta > \pi/4$.

\begin{figure}
\includegraphics[width=88mm]{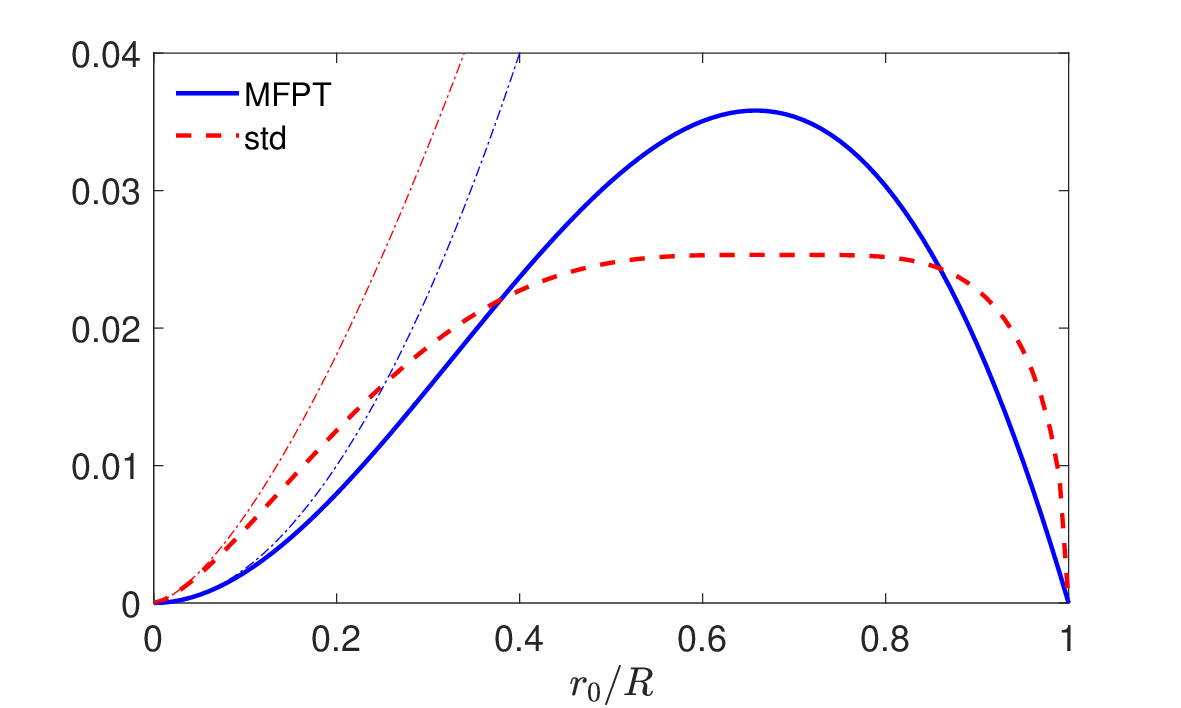} 
\caption{
MFPT (solid line) and the standard deviation of the FPT (dashed line)
as functions of the distance $r_0$ to the origin of the sector
$\Omega_{\Theta,R}$, with $\Theta = \pi/3$ and $R = 1$.  Thin
dash-dotted lines present the asymptotic relations
(\ref{eq:TTheta_asympt}, \ref{eq:std_asympt}).}
\label{fig:MFPT_std}
\end{figure}

Figure \ref{fig:MFPT_std} illustrates the behavior of the MFPT and the
standard deviation of the FPT for the sector with $\Theta = \pi/3$.
When the starting point is far from the boundary ($r_0/R \sim 0.5$),
the standard deviation is comparable to the MFPT that is a common
situation for diffusive search in a bounded domain.  In turn, as $r_0
\to 0$, both MFPT and the standard deviation tend to $0$, but the MFPT
vanishes much faster.  As a consequence, the standard deviation is
much bigger than the MFPT, while their ratio diverges according to
Eq. (\ref{eq:coeff_variation}).

\end{document}